\documentclass[12pt,epsfig,color]{article}
\usepackage{amsmath}
\usepackage{epsfig}
\usepackage{amssymb}
\input{epsf}
\newcommand\as{\alpha_s}
\newcommand\f[2]{\frac{#1}{#2}}
\def\la{\lambda} 
 
\def\beq{\begin{equation}}
\def\eeq{\end{equation}}
\def\beeq{\begin{eqnarray}}
\def\eeeq{\end{eqnarray}}
\def\to{\rightarrow}
\def\nn{\nonumber}

\def\msbar{{\overline {\rm MS}}} 
 
\def\b0{b_0}
\def\bone{b_1}

\begin{document}

\begin{titlepage}
\renewcommand{\thefootnote}{\fnsymbol{footnote}}
\begin{flushright}
BNL-NT-05/9 \\
HUPD-0503 \\
RBRC-487 \\  
YITP-SB-05-07 \\
\today
     \end{flushright}
\par \vspace{7mm}
\begin{center}
{\Large \bf
Dilepton production near partonic threshold \\[4mm]
in transversely polarized $\bar{p}p$ collisions}

\end{center}
\par \vspace{3mm}
\begin{center}
\large 
{Hirotaka\ Shimizu$^{a}$, 
George\ Sterman$^{b}$, \\[2mm]
Werner\ Vogelsang$^{c}$, and 
Hiroshi\ Yokoya$^{a,c}$}

\vspace*{0.5cm}
\normalsize
{\em $^a$Department of Physics, Hiroshima University, 
Higashi-Hiroshima 739-8526, Japan} \\

\vspace*{0.5cm}
{\em $^b$C.N.\ Yang Institute for Theoretical Physics,
Stony Brook University \\
Stony Brook, New York 11794 -- 3840, U.S.A.} \\

\vspace*{0.5cm}
{\em $^c$Physics Department and RIKEN-BNL Research Center, 
Brookhaven National Laboratory,\\
Upton, New York 11973, U.S.A.}\\

\end{center}

\par \vspace{5mm}
\begin{center} {\large \bf Abstract} \end{center}
\begin{quote}
\pretolerance 10000
It has recently been suggested that collisions of 
transversely polarized protons and antiprotons at the GSI 
could be used to determine the nucleon's transversity densities
from measurements of the double-spin asymmetry for the Drell-Yan 
process. We analyze the role of higher-order perturbative QCD 
corrections in this kinematic regime, in terms of the available
fixed-order contributions as well as of all-order soft-gluon
resummations. We find that the combined perturbative corrections to 
the individual unpolarized and transversely polarized cross sections 
are large. 
We trace these large enhancements to soft gluon emission near 
partonic threshold, and we suggest that with a physically-motivated 
cut-off enhancements beyond lowest order are moderated relative to 
resummed perturbation theory, but still significant. The unpolarized 
dilepton cross section for the GSI kinematics may therefore provide 
information on the relation of perturbative and 
nonperturbative dynamics in hadronic scattering. The spin asymmetry 
turns out to be rather robust, relatively insensitive to higher 
orders, resummation, and the cut-offs.

\end{quote}

\end{titlepage}

\setcounter{footnote}{1}
\renewcommand{\thefootnote}{\fnsymbol{footnote}}

\section{Introduction}
\noindent
The partonic structure of polarized nucleons 
at the leading-twist level  is
characterized by the unpolarized, 
longitudinally polarized, and transversely polarized parton 
distribution functions $f$, $\Delta f$, and $\delta f$, 
respectively \cite{ref:jaffeji}. 
In contrast to the distributions $f$ and $\Delta f$, 
we have essentially no knowledge from experiment so far about the 
transversity distributions $\delta f$, even though there are
now first indications~\cite{hermest} that some of them 
are non-vanishing. The $\delta f$ were
first introduced in \cite{ref:ralston}. They are defined
as \cite{ref:jaffeji,ref:ralston,ref:artru,ref:ratcliffe} the difference of 
probabilities for finding a parton of flavor $f$ at scale $\mu$ and
light-cone momentum fraction $x$ with its spin aligned 
($\uparrow\uparrow$) or anti-aligned ($\downarrow\uparrow$)
to that of the transversely polarized nucleon:
\begin{equation}
\label{eq:pdf}
\delta f(x,\mu) \equiv f_{\uparrow\uparrow}(x,\mu) -
                       f_{\downarrow\uparrow}(x,\mu) \; .
\end{equation}

By virtue of factorization theorems \cite{ref:fact1,ref:fact2}, the 
parton densities can be probed universally in a variety of
inelastic scattering processes for which it is possible
to separate (``factorize'') the long-distance physics relating
to nucleon structure from a partonic short-distance scattering
that can be calculated in QCD perturbation theory.
It was realized a long time ago \cite{ref:jaffeji,ref:ralston,ref:artru}
that due to its chirally-odd structure, transversity decouples from 
inclusive deeply-inelastic scattering, but that inelastic
collisions of two transversely polarized nucleons should offer
good possibilities to access transversity. In particular, 
the Drell-Yan processes $pp\to l^+l^-X$, $p\bar{p}\to l^+l^-X$ $(l=e,\mu)$
were identified as promising sources of information on 
transversity~\cite{ref:dylo,ref:dy1,ref:dy1a}. 
This is so because there is no gluon transversity distribution at leading 
twist~\cite{ref:jaffeji,ref:artru}. For the Drell-Yan process, the 
lowest-order partonic process is $q\bar{q}\to \gamma^{\ast}$,
with gluonic contributions to the unpolarized cross
section in the denominator of the transverse double-spin asymmetry
\begin{equation} \label{attdefi}
A_{TT}=\frac{\sigma^{\uparrow\uparrow}-\sigma^{\uparrow\downarrow}}
{\sigma^{\uparrow\uparrow}+\sigma^{\uparrow\downarrow}}
\end{equation}
only arising as higher-order corrections. Therefore, $A_{TT}$ may 
be sizable for the Drell-Yan process, 
in contrast to other hadronic processes such as 
high-transverse-momentum prompt photon and jet 
production~\cite{ref:artru,ref:attold,ref:jaffesaito,ref:attlo,ref:attnlo}
which in the unpolarized case are largely driven by gluons in the initial
state and are hence expected to have a very suppressed $A_{TT}$. 

Clean information on transversity should be gathered from polarized 
proton-proton collisions at the BNL Relativistic Heavy Ion Collider 
(RHIC) where the Drell-Yan process is a major focus~\cite{review}. 
In $pp$ collisions, however, the Drell-Yan process probes products of
valence quark and sea antiquark distributions. It is possible that 
antiquarks in the nucleon carry only little transverse polarization 
since, due to the absence of a 
gluon transversity, a source for the perturbative generation of 
transversity sea quarks from $g\to q\bar{q}$ splitting is missing. 
In addition, at RHIC energies and for Drell-Yan masses of a 
few GeV, the partonic momentum fractions are fairly small, so that
the denominator of $A_{TT}$ is large due to the small-$x$ rise
of the unpolarized sea quark distributions. Thus, even for the 
Drell-Yan process, the spin asymmetry $A_{TT}$ at RHIC will
probably be at most a few per cent, as theoretical 
studies have shown \cite{ref:dy1,ref:dy1a,ref:kkst}. 

It has recently been proposed to add
polarization to planned $\bar{p}p$ collision experiments 
at the GSI, and to perform measurements of $A_{TT}$ for the Drell-Yan 
process~\cite{dypax1,dypax1a,dypax2,dypax3}. 
This is a very exciting idea, since
unique information on transversity could be obtained in this way. 
The results would be complementary 
to what can be obtained from RHIC measurements. First of all, in $\bar{p}p$
collisions the Drell-Yan process mainly probes products 
of two {\it quark} densities, $\delta q\times \delta q$,
since the distribution of antiquarks in antiprotons equals
that of quarks in the proton. In addition, kinematics in 
the proposed experiments are such that rather large partonic
momentum fractions, $x\sim 0.5$, are probed. One therefore
accesses the valence region of the nucleon. 
Estimates \cite{dypax2,dypax3,dypax4}
for the GSI PAX and ASSIA experiments show that the expected spin 
asymmetry $A_{TT}$ should be very large, of order 40\% or more. 

It is important, however, to keep in mind that the kinematic
region to be accessed in the first stage of the GSI experiments, with 
Drell-Yan masses $M$ of $1-4$~GeV or so, but a center-of-mass
energy of only $\sqrt{S}\approx 5.3$~GeV for the baseline fixed-target
program, is not really
the ``classic'' regime where parton model ideas, factorization, 
and perturbative QCD are a priori expected to provide adequate 
descriptions. This is of course crucial since the interpretation 
of $A_{TT}$ in terms of transversity relies exactly on these concepts.
To be more precise, at high energies and large dilepton invariant mass $M$
the cross section factorizes \cite{ref:fact1,ref:fact2} into 
convolutions of parton densities and perturbative partonic 
hard-scattering cross sections, as mentioned above. 
Schematically,
\begin{equation}\label{fact}
M^4\,\frac{d\sigma}{dM^2}=\sum_{a,b}\,
f_a\otimes f_b\otimes \frac{M^4d\hat{\sigma}_{ab}}{dM^2}\;+\;
{\cal O}\left(  \frac{\lambda}{M}\right)^p \; .
\end{equation} 
For simplicity, we have considered here the unpolarized cross 
section, and we have also integrated over the rapidity of the lepton
pair and only focused on the total Drell-Yan cross section.
We also have not written out the precise form of the convolutions,
which will be given below. For the moment, we are only interested
in the important features visible in Eq.~(\ref{fact}). The 
quantities one wants to determine from measurement of the 
left-hand-side of Eq.~(\ref{fact}) are the parton distributions
$f_a,f_b$. The partonic cross sections, $\hat{\sigma}_{ab}$,
for the reactions $ab\to \gamma^{\ast}X$ may be calculated in
QCD perturbation theory. Their expansion in terms of the 
strong coupling constant $\alpha_s(M)$ reads  
\begin{equation}
d\hat{\sigma}_{ab}=d\hat{\sigma}_{ab}^{(0)}+\frac{\alpha_s(M)}{\pi}
d\hat{\sigma}_{ab}^{(1)}+\left( \frac{\alpha_s(M)}{\pi}\right)^2
d\hat{\sigma}_{ab}^{(2)}+\ldots \; ,
\end{equation}
corresponding to lowest order (LO), next-to-leading order (NLO), 
and so forth. The earlier studies \cite{dypax2,dypax3} for the 
Drell-Yan process at GSI energies used LO hard-scattering 
cross sections to estimate the expected spin asymmetries. 
Depending on kinematics, however, the higher-order corrections 
may be very important. As we will show below, this is the case
for the planned GSI measurements. In addition, as indicated 
in Eq.~(\ref{fact}), factorization of the hadronic cross
section in terms of twist-2 distributions 
is of course not exact, but holds only to leading power
in $M$. There are corrections to the (dimensionless) cross 
section $M^4d\sigma/dM^2$ that are down by inverse powers of the 
hard scale, that is, of the form $\left(  \lambda/M\right)^p$
with some $p$ and some hadronic mass scale $\lambda$~\cite{JG}. 
These power corrections will also depend on $\tau=M^2/S$ and
are generally expected to increase with increasing $\tau$.
The measured spin asymmetry $A_{TT}$ can only be reliably
interpreted in terms of the transversity densities
if the higher order and power corrections can either be shown to be 
small in the accessible kinematic domain, and/or if they are 
sufficiently well understood.
The aim of this paper is to address primarily the question
of how large the higher-order QCD corrections to the 
Drell-Yan process are in the GSI kinematic regime. 
To this end, we will apply the technique of threshold resummation, 
to which we now turn. 

As we mentioned above, $\tau$ is typically very large
for the GSI kinematics, $0.2 \lesssim \tau\lesssim 0.7$. 
This is a region where higher-order corrections to the partonic 
cross sections are particularly important. $\tau=1$ sets a 
threshold for the reaction, and as $\tau$ increases toward unity,
very little phase space for real gluon radiation remains in 
the partonic process, since most of the initial partonic energy 
is used to produce the virtual photon. Virtual and real-emission
diagrams then become strongly imbalanced, and the infrared cancellations
leave behind large logarithmic higher order corrections to the 
partonic cross sections, the so-called threshold logarithms. 
At the $k$th order in perturbation theory, the leading logarithms
are of the form $\as^k\ln^{2k-1}(1-z)/(1-z)$, where
$z=\tau/x_a x_b$ is the partonic analogue of $\tau$. 
For sufficiently large $z$, perturbative calculations to fixed 
order in $\alpha_s$ become unreliable, since the double logarithms 
compensate the smallness of $\alpha_s(M)$ even if $M$ is of the 
order of a few GeV. The fact that the parton distributions are 
steeply falling functions of the momentum fractions $x_{a,b}$ 
means that the threshold region is actually emphasized in the cross 
section, even if $\tau$ itself is still rather far away from one. 
If $\tau$ is close to unity, as is the case for much of the 
GSI kinematics, the region of large $z\lesssim 1$ completely 
dominates, and it is crucial that the terms $\as^k\ln^{2k-1}(1-z)/(1-z)$
be resummed to all orders in $\alpha_s$. Such a ``threshold resummation'' 
was originally developed for the Drell-Yan process~\cite{dyresum1,dyresum2}
and subsequently applied to a variety of more involved 
partonic processes in QCD~\footnote{See, for example, the review
in Ref.~\cite{Kidonakis}, and interesting recent
work~\cite{IJ} that rederives some of these results in the 
context of the soft-collinear effective theory~\cite{SCET}.}.
It turns out that the soft-gluon
effects exponentiate, not in $z$-space directly, but in
Mellin-$N$ moment space, where $N$ is the Mellin moment conjugate
to $z$. The leading logarithms (LL) in the exponent are of the form 
$\as^k \ln^{k+1} (N)$, subleading logarithms (next-to-leading 
logarithms (NLL)) of $\as^k\ln^k (N)$. In this paper we will
use NLL resummed perturbation theory to analyze the importance
of higher-order corrections to the Drell-Yan cross section 
in the kinematic regime to be explored at the GSI. 

There is a close relation between resummation and the 
nonperturbative power corrections. Taking into account 
nonleading logarithms and the running of the coupling, 
resummation always leads to a perturbative expression in 
which the scale of the coupling reflects the value of the 
transform variable. Because of the singularity of the perturbative 
effective coupling at 
$\Lambda_{\mathrm QCD}$, the resulting expressions are 
ill-defined~\cite{renormalon}. The analysis of these ambiguities 
for Drell-Yan cross sections~\cite{cspv,KSV} suggests a 
series of nonperturbative corrections~\cite{cspv}, generically 
suppressed by even powers of the pair mass $M$, but enhanced by 
the moment variable $N$, $N^2/M^2$. As we shall see, for GSI
fixed-target energies, the effective values of $N$ are so large that 
the first few power 
corrections will not suffice for the Drell-Yan cross section. 
We will therefore rely on a somewhat different approach, to be 
presented in more detail elsewhere, and cut off unphysical 
dependence on low momentum scales. The result for a large 
portion of dilepton masses $M$ will be a cross section with 
moderated, but still significant enhancements relative even 
to next-to-next-to-leading order calculations, which we take as a 
conservative prediction based on perturbation theory. Experimental 
results on these cross sections should shed light on the 
interrelations between fixed-order, all-order and nonperturbative 
corrections in hadronic scattering. 

The remainder of this paper is organized as follows. Section~\ref{sec2}
will present the basic framework for our calculations and 
will introduce the partonic threshold region. In sec.~\ref{sec3}, we
provide all ingredients for the NLL resummation of 
the threshold logarithms, and we also propose a new infrared-regulated 
expression for the form of nonperturbative corrections suggested by
perturbative resummation. In sec.~\ref{sec4} 
we then present phenomenological results for the
regions of interest in GSI measurements.  

\section{Perturbative cross section and the threshold region
\label{sec2}}

The spin-dependent cross section for dilepton production
by two transversely polarized hadrons is defined as
\begin{equation} \label{wqdef}
\frac{\tau d\delta \sigma}{d\tau d\phi} \equiv
\frac{1}{2} \left( \frac{\tau d\sigma^{\uparrow \uparrow}}{d \tau
d\phi} - \frac{\tau d\sigma^{\uparrow \downarrow}}{d\tau d\phi}
\right) \; ,
\end{equation}
where the superscript $\uparrow \uparrow$ ($\uparrow \downarrow$)
denotes parallel (antiparallel) setting of the transverse spins
of the incoming hadrons. We have used the customary Drell-Yan
scaling variable $\tau=M^2/S$ with $M$ the invariant mass 
of the lepton pair and $S$ the center-of-mass energy squared. 
$\phi$ is the azimuthal angle of one of the leptons, 
counted relative to the axis defined by the transverse 
polarizations. For simplicity, we have integrated over all 
rapidities of the lepton pair. At high $M$, the cross section factorizes
into convolutions of the transversity distributions $\delta f$
with the corresponding transversely polarized partonic 
hard-scattering cross sections \cite{ref:fact2}:
\begin{equation} 
\label{sig} 
\frac{\tau d\delta\sigma(\tau)}{d\tau d\phi}=\sum_{a,b}
\sigma_{ab}^{(0)}(\phi) 
\int_{\tau}^1 \frac{dx_a}{x_a} \int_{\tau/x_a}^1 \frac{dx_b}{x_b} \;
\delta f_a(x_a,\mu^2)\, \delta f_b(x_b,\mu^2)\;
\delta \omega_{ab}\left(z\equiv\tau/x_a x_b,\frac{M^2}{\mu^2},
\alpha_s(\mu) \right) \;,
\end{equation}
where $\mu$ collectively denotes the factorization and renormalization 
scales, and where we will specify the $\sigma_{ab}^{(0)}(\phi)$ below.
Due to the odd chirality of transversity, and since there is no 
transversity gluon density, $q\bar{q}$ annihilation is the only 
partonic channel, up to next-to-leading order (NLO)\footnote{Starting 
from next-to-next-to-leading order (NNLO) there are contributions 
from $qq$ scattering as well.}. Its cross section is calculated in 
QCD perturbation theory as a series in $\alpha_s$:
\begin{equation}
\delta \omega_{ab}\left(z,r,\alpha_s \right)= \delta \omega_{ab}^{(0)}
\left(z\right)+
\frac{\alpha_s}{\pi}  \delta \omega_{ab}^{(1)}\left(z,r\right)+
\left(\frac{\alpha_s}{\pi}\right)^2 
\delta \omega_{ab}^{(2)}\left(z,r\right)+\ldots \; ,
\end{equation}
where $r=M^2/\mu^2$. Since we will not consider very high energies and 
dilepton masses, only photons contribute as intermediate particles.
The lowest-order (LO, ${\cal O}(\alpha_s^0)$) process thus is
$q^{\uparrow}\bar{q}^{\uparrow}\to \gamma^{\ast}\to l^+ l^-$, for 
which
\begin{equation}
\label{eq:lo}
\delta \omega_{q\bar{q}}^{(0)}\left(z\right)=
\delta(1-z) \; \; , \;\;\;\; 
\sigma_{q\bar{q}}^{(0)}(\phi)=\frac{\alpha^2e_q^2}{9S} 
\cos (2\phi)
\; .
\end{equation}
The first-order term $\delta \omega^{(1)}_{q\bar{q}}$ is known and reads \cite{ref:dy2}
\begin{eqnarray} \label{cq1z}
\delta \omega_{q\bar{q}}^{(1)} (z,r) &=& C_F \left[ 4z \left( 
\frac{\ln (1-z)}{1-z} \right)_+ - \frac{2 z \ln z}{1-z}-\frac{3 z 
\ln^2 z}{1-z} + 2 (1-z) \right. \nonumber \\
&& \hspace*{0.9cm} + \left( \frac{\pi^2}{3}  -4 \right) \delta (1-z) 
 + \left( \frac{2z}{(1-z)_+}+\frac{3}{2}\delta(1-z) \right)\ln r 
\Bigg] \; ,
\end{eqnarray}
where $C_F=4/3$ and the ``+''-distribution is defined as 
\begin{equation}
\int_0^1 dz \,\left[g(z)\right]_+ \, f(z) =
\int_0^1 dz \,g(z) \,\left(f(z)-f(1) \right) \; .
\end{equation}

Since resummation is performed in Mellin moment space, we take a 
Mellin transform of the hadronic cross section:
\begin{equation} \label{mellin}
\frac{d\delta \sigma^N}{d\phi} \equiv \int_0^1 d\tau \tau^{N-1}
\frac{\tau d\delta \sigma}{d \tau d\phi} \; .
\end{equation}
The cross section algebraically factorizes under moments,
\begin{equation} \label{nlogen}
\frac{d\delta \sigma^N}{d\phi} = \sigma_0\sum_q\,
\delta q^N(\mu^2)\,
\delta \bar{q}^N(\mu^2)\;
\left[ 1 + \frac{\alpha_s (\mu)}{\pi} \delta \omega_{q\bar{q}}^{(1),N}(r)+ 
\left( \frac{\alpha_s (\mu)}{\pi}\right)^2 
\delta \omega_{q\bar{q}}^{(2),N}(r)+\ldots
\right] \; ,
\end{equation}
where the Mellin moments of the transversity 
distributions $\delta q$ and the higher-order corrections 
$\delta \omega_{q\bar{q}}^{(k)}$ are defined as usual,
\begin{eqnarray}
\delta q^N (\mu^2) &=& \int_0^1 dx x^{N-1} 
\delta f_q(x,\mu^2) \; , \nonumber \\
\delta \omega_{q\bar{q}}^{(k),N}(r)&=& \int_0^1 dz z^{N-1}
\delta \omega_{q\bar{q}}^{(k)}(z,r) \; .
\end{eqnarray} 
The moments of the first-order correction 
$\delta \omega_{q\bar{q}}^{(1)}$ in Eq.~(\ref{cq1z})
read in the $\overline{\mathrm{MS}}$ scheme~\cite{ref:dy2}:
\begin{equation} \label{fullnlo}
\delta \omega_{q\bar{q}}^{(1),N}(r) =  C_F \left[ \frac{2}{N(N+1)}
+ 2 S_1^2 (N) + 6 \left( S_3 (N) - \zeta (3) \right) - 4 + 
\frac{2}{3} \pi^2 + \left( \frac{3}{2} - 2 S_1(N) \right)\ln r \right] \; .
\end{equation} 
The sums appearing here are defined by
\begin{eqnarray}
S_k (N) &\equiv& \sum_{j=1}^N \frac{1}{j^k}  \; .
\end{eqnarray}
Their analytic continuations to arbitrary Mellin-$N$ are
\begin{eqnarray}
S_1(N)&=&\psi(N+1)+\gamma_E\; , \nonumber\\
S_3(N)&=& \frac{1}{2}\psi''(N+1)+\zeta(3)\; , 
\end{eqnarray}
where $\psi(z)$ is the digamma function, $\gamma_E=0.5772\ldots$ is 
the Euler constant, and $\zeta(3)\approx 1.202057$. 

We mention that formulas analogous to the above hold for the 
unpolarized case. The main difference is that in the unpolarized case 
beyond LO there are contributions from initial-state gluons to the Drell-Yan 
cross section. In the kinematic region we are interested in here, these are
rather unimportant. All details for the unpolarized case to NLO
may for example be found in Ref.~\cite{ref:dy1}.

The threshold region corresponds to $z\to 1$ or $N\to \infty$.
At large $N$, the moments of the NLO correction become
\begin{equation} \label{c1qlargen}
\delta \omega_{q\bar{q}}^{(1),N}(r) =  C_F \left[ 
2 \ln^2 (\bar{N})  - 4 + 
\frac{2}{3} \pi^2 + \left( \frac{3}{2} - 2 \ln(\bar{N}) \right)\ln r \right] +
{\cal O}\left( \frac{1}{N}\right)\; ,
\end{equation} 
where 
\beq
\bar{N}=N {\rm e}^{\gamma_E} \; .
\eeq
One can see the double-logarithmic corrections $\propto
\as\ln^2 (\bar{N})$ near threshold, associated with the 
logarithmic term $\sim \ln(1-z)/(1-z)$ in Eq.~(\ref{cq1z}).
At higher orders, there are corrections of the form $\as^k\ln^{l}(\bar{N})$
with $l\leq 2k$.  We emphasize that the behavior of the unpolarized 
partonic cross section near threshold is exactly the same
as Eq.~(\ref{c1qlargen}). This is related to the fact that 
the large logarithms are due to the emission of soft gluons, which 
is spin-independent. Thanks to the simple structure of the LO Drell-Yan 
process, the constant ($N$-independent) pieces which are partly associated 
with virtual corrections are identical as well for the transversely 
polarized and unpolarized cases, provided both are treated in the same
factorization scheme.

We now turn to the resummation of the leading and next-to-leading
threshold logarithms to all orders in $\alpha_s$. 

\section{Resummed cross section \label{sec3}}

In Mellin-moment space, threshold resummation for the 
Drell-Yan process results in the exponentiation of the soft-gluon
corrections. To NLL~\footnote{We note that the threshold resummation for the 
Drell-Yan process has been worked out even to next-to-next-to-leading 
logarithmic accuracy, see~\cite{vogtnnll}.}, the resummed formula 
is given in the $\msbar$ scheme by
\begin{equation} \label{dyres}
\delta \omega_{q\bar{q}}^{{\rm res},N}(r,\as(\mu))=
\exp\left[ C_q (r,\as(\mu)) \right]
\exp\left\{ 2 \int_0^1 dz\, \f{z^{N-1}-1}{1-z} 
\int_{\mu^2}^{(1-z)^2 M^2} \f{dk_T^2}{k_T^2} A_q(\as(k_T))\right\} \; ,
\end{equation}
where 
\begin{equation} \label{andim}
A_q(\as)=\frac{\as}{\pi} A_q^{(1)} +  
\left( \frac{\as}{\pi}\right)^2 A_q^{(2)} + \ldots \; ,
\end{equation}
with~\cite{KT}:
\begin{equation} 
\label{A12coef} 
A_q^{(1)}= C_F
\;,\;\;\;\; A_q^{(2)}=\frac{1}{2} \; C_F  \left[ 
C_A \left( \frac{67}{18} - \frac{\pi^2}{6} \right)  
- \frac{5}{9} N_f \right] \; ,
\end{equation}  
where $N_f$ is the number of flavors and $C_A=3$. 
The coefficient $C_q (r,\as(\mu))$ collects mostly 
hard virtual corrections. It is a perturbative 
series and reads
\begin{eqnarray} 
C_q (r,\as(\mu))=\frac{\as}{\pi}\,C_F\,
\left( -4+\frac{2\pi^2}{3}  +\frac{3}{2}\,\ln r \right)
 + {\cal O}(\as^2) \; .
\end{eqnarray} 
We note that it was shown in~\cite{Eynck:2003fn} that these corrections
also exponentiate. 

Eq.~(\ref{dyres}) as it stands is ill-defined because
of the divergence in the perturbative running coupling 
$\alpha_s(k_T)$ at $k_T=\Lambda_{\rm QCD}$. The perturbative 
expansion of the expression shows factorial divergence, 
which in QCD corresponds to a power-like ambiguity of the series.
It turns out, however, that the factorial divergence appears only
at nonleading powers of momentum transfer. The large logarithms
we are resumming arise in the region~\cite{dyresum2} 
$z\leq 1-1/\bar{N}$ in the integrand in Eq.~(\ref{dyres}). One therefore 
finds that to NLL they are contained in the simpler expression
\begin{equation} \label{dyres1}
2 \int_{M^2/\bar{N}^2}^{M^2} \f{dk_T^2}{k_T^2} A_q(\as(k_T))
\ln\frac{\bar{N}k_T}{M}\,+\,
2 \int_{M^2}^{\mu^2} \f{dk_T^2}{k_T^2} A_q(\as(k_T))
\ln\bar{N}
\end{equation}
for the second exponent in~(\ref{dyres}).
This form, to which we will return below, is used for ``minimal'' 
expansions~\cite{Catani:1996yz} of the resummed exponent. 

\subsection{Exponents at NLL}

In the exponents, the large logarithms in $N$
now occur only as single logarithms, of the form 
$\as^k \ln^{k+1}(N)$ for the leading terms. Subleading terms 
are down by one or more powers of $\ln(N)$. Knowledge of the 
coefficients $A_q^{(1,2)}$ in Eq.~(\ref{dyres}) 
is enough to resum the full towers of LL terms
$\as^k \ln^{k+1}(N)$, and NLL ones $\as^k \ln^k(N)$ in the exponent. 
With the coefficient $C_q$ one then gains 
control of three towers of logarithms in the cross section, 
$\as^k \ln^{2k}(N)$, $\as^k \ln^{2k-1}(N)$, $\as^k \ln^{2k-2}(N)$.

We now give the explicit formula for the expansion of the 
resummed exponent to NLL accuracy. From Eqs.~(\ref{dyres}),(\ref{dyres1})
one finds \cite{Catani:1996yz,CMN}
\beeq
\label{lndeltams}
\ln \delta \omega_{q\bar{q}}^{{\rm res},N} (r,\as(\mu)) 
&=& C_q (r,\as(\mu))+2\ln \bar{N} \;h^{(1)}(\lambda) +
2 h^{(2)}(\lambda,r) \; ,
\eeeq
where 
\beq  \label{lamdef}
\lambda=\b0 \as(\mu) \ln \bar{N} \; . 
\eeq
The functions $h^{(1,2)}$ are given by 
\begin{align} 
\label{h1fun}
h^{(1)}(\la) =& \f{A_q^{(1)}}{2\pi \b0 \la} 
\left[ 2 \la+(1-2 \la)\ln(1-2\la)\right] \;,\\ 
h^{(2)}(\la,r) 
=&-\f{A_q^{(2)}}{2\pi^2 \b0^2 } \left[ 2 \la+\ln(1-2\la)\right]
+ \f{A_q^{(1)} \bone}{2\pi \b0^3} 
\left[2 \la+\ln(1-2\la)+\f{1}{2} \ln^2(1-2\la)\right]\nn \\ 
\label{h2fun}
&+ \f{A_q^{(1)}}{2\pi \b0}\left[2 \la+\ln(1-2\la) \right]  
\ln(r)-\f{A_q^{(1)}\as(\mu)}{\pi} \,\ln \bar{N}\, \ln(r) \;,  
\end{align} 
where
\begin{eqnarray}
\b0 &=& \frac{1}{12\pi} \left( 11 C_A - 2 N_f \right) \; , \nn \\
\bone&=&  \frac{1}{24 \pi^2} 
\left( 17 C_A^2 - 5 C_A N_f - 3 C_F N_f \right) \;\; .
\label{bcoef}
\end{eqnarray}
The function $h^{(1)}$ contains all LL terms in the perturbative 
series, while $h^{(2)}$ is of NLL only. 
We note that the resummed exponent depends on the 
factorization scales in such a way
that it will compensate the scale dependence (evolution) 
of the parton distributions. This feature is represented by 
the last term in~(\ref{h2fun}). One therefore expects a decrease in 
scale dependence of the cross section from resummation. 
The remaining $\mu$-dependence in the second to last term
in~(\ref{h2fun}) results from writing the strong coupling constant as
\begin{equation}
\alpha_s(k_T)=\frac{\alpha_s(\mu)}{1+b_0 \alpha_s(\mu)
\ln (k_T^2/\mu^2)}+{\cal O}\left( \alpha_s(\mu)^2 
(\alpha_s(\mu)\ln (k_T^2/\mu^2))^n\right)
\end{equation}
when doing the NLL expansion of the exponent. For this term, $\mu$ 
represents the renormalization scale.

As was shown in Refs.~\cite{KSV,cat}, it is possible to improve 
the above formula slightly and to also correctly take into account certain
subleading terms in the resummation. To this end, we rewrite
Eqs.~(\ref{lndeltams})-(\ref{h2fun}) as 
\beeq
\label{lndeltams1}
\!\!\! \!\!\! \!\!\! \!\!\! \!\!\!
\ln \delta \omega_{q\bar{q}}^{{\rm res},N} (r,\as(\mu)) 
&=& 
\frac{1}{\pi b_0}
\left[ 2 \la+\ln(1-2\la)\right]\,\left( \frac{A_q^{(1)}}{b_0 \alpha_s(\mu)}-
\frac{A_q^{(2)}}{\pi b_0}+\frac{A_q^{(1)}b_1}{b_0^2}+
A_q^{(1)}\,\ln r
\right)\nonumber \\
&+&
\frac{\as}{\pi}\,C_F\,
\left( -4+\frac{2\pi^2}{3}  \right)+
\frac{A_q^{(1)}b_1}{2\pi b_0^3} \,\ln^2(1-2\la)+
B_q^{(1)}\,\frac{\ln(1-2 \la)}{\pi b_0}
\nonumber \\
&+&\left[-2 A_q^{(1)}\ln \bar{N}-B_q^{(1)}\right]\,
\left( \frac{\alpha_s(\mu)}{\pi} \ln r + \frac{\ln(1-2 \la)}{\pi b_0}
\right)\; ,
\eeeq
where $B_q^{(1)}=-3 C_F/2$. The last term in Eq.~(\ref{lndeltams1})
is the LL expansion of the term 
\beq\label{evol}
\int_{\mu^2}^{M^2/\bar{N}^2} \f{dk_T^2}{k_T^2} \,\frac{\as(k_T)}{\pi} \,
\left[-2 A_q^{(1)}\ln \bar{N}-B_q^{(1)}\right]\; .
\eeq
Since the factor $\left[-2 A_q^{(1)}\ln \bar{N}-B_q^{(1)}\right]$ 
is the large-$N$
limit of the moments of the LO DGLAP splitting function for transversity,
$\delta P^N$, the term in Eq.~(\ref{evol}) may be viewed as an 
evolution of the parton distributions between scales $\mu$ and 
$M/\bar{N}$. This suggests to modify the resummation by 
replacing~\cite{KSV,Kra98}
\beq\label{evol1}
\left[-2 A_q^{(1)}\ln \bar{N}-B_q^{(1)}\right]\;\;\longrightarrow \;\;
\delta P^N\equiv C_F \left[ \frac{3}{2}-2 S_1(N)\right]
\eeq
in Eq.~(\ref{lndeltams1}). To see the improvement resulting from this,
we expand the resummed formula in Eq.~(\ref{lndeltams1}), after
the replacement~(\ref{evol1}), to 
first order in $\alpha_s(\mu)$ and find:
\beq
\delta \omega_{q\bar{q}}^{{\rm res},N} (r,\as) = 1 + \frac{\alpha_s}{\pi} C_F
\left[4 \ln(\bar{N})S_1(N)-2 \ln^2(\bar{N}) - 4 + 
\frac{2}{3} \pi^2 +
\left( \frac{3}{2} - 2 S_1(N) \right)\ln r
\right]+{\cal O}(\as^2)\; .
\eeq
This term correctly gives the large-$N$ pieces of the 
NLO cross section in Eq.~(\ref{c1qlargen}), but it goes beyond
that by also reproducing all contributions $\sim \ln(\bar{N})/N$
in Eq.~(\ref{fullnlo}), the latter arising from the expansion
$S_1(N)=\ln(\bar{N})+1/(2N)+{\cal O}(1/N^2)$. 

The unpolarized resummed partonic cross section in Mellin-moment space 
is practically identical to the transversely polarized one 
in Eq.~(\ref{dyres}), since the coefficients $A(\as)$ and 
$C_q(r,\as)$ are spin-independent and thus the same for the unpolarized
and transversely polarized cases. A very small difference 
arises in the replacement in Eq.~(\ref{evol1}), in which for the 
unpolarized case one of course has to use the unpolarized LO splitting 
function. In addition, one should also take into account the singlet 
mixing in the evolution~\cite{KSV} which, however, is very small in the 
kinematic region we are interested in.

\subsection{Far-infrared resummed cross section}

We will see in the phenomenology section below that perturbative 
resummation as formulated so far in Eqs.\ (\ref{dyres}) or (\ref{dyres1})
predicts very large enhancements of the lowest order cross section,
sometimes by orders of magnitude.  There is good reason 
to believe that this enhancement is only partly physical. The large 
corrections arise from a region where the integral in the 
exponent becomes sensitive to the behavior of the integrand at small 
values of $k_T$. As long as $\Lambda_{\mathrm QCD}\ll M/
\bar{N}\ll M$, the use of NLL perturbation theory may be justified, but 
when $|N|$ becomes so large that $k_T$ goes down to nonperturbative scales,  
we may well question the self-consistency of perturbation theory.  We now 
turn to  the question of how this ``far-infrared" limit should be treated 
in resummed perturbation theory. Our discussion here will be brief, 
and we will only present one model that addresses the far-infrared limit.
A more detailed study of this very interesting regime will be presented  
in a future publication. 

We seek a modification of the perturbative 
expression Eq.\ (\ref{dyres}) that excludes the region
in which the absolute value of $k_T$ is less than some scale
$\mu_0> m_\pi$.  We think of $\mu_0$ as the scale beyond which
the true mass spectrum of QCD replaces perturbation theory,
regulating all soft and collinear singularities, so that $m_\pi$ should
be thought of only as a lower limit for $\mu_0$. To implement this
idea, we will adopt a modified resummed hard scattering, which
reproduces NLL logarithmic behavior in the moment variable $N$
so long as $M/\bar{N}>\mu_0$, but ``freezes"  once 
$M/\bar{N}<\mu_0$.  If nothing else, this will test the importance
of the region $k_T \le \Lambda_{\mathrm  QCD}$ for the resummed cross
section. If $N$ were real and positive, 
we could simply replace the first exponent 
in~(\ref{dyres1}) by
\begin{equation} \label{dyres2}
4 \int_{\rho(M/\bar{N},\, \mu_0)}^M \f{dk_T}{k_T} A_q(\as(k_T))
\ln\frac{\bar{N}k_T}{M} \; ,
\end{equation}
where 
\begin{equation} \label{dyres3}
\rho(a,b)=\max(a,b) \; ,
\end{equation}
and where $\mu_0$ then serves to cut off the lower logarithmic behavior.
To provide an expression that can be
continued to complex $N$, we choose 
\begin{equation} \label{dyres4}
\rho(a,b)=(a^p+b^p)^{1/p} \; ,
\end{equation}
with integer $p$. This simple form is consistent with the minimal 
expansion given above, and it also allows for
a straightforward analysis of the ensuing branch cuts in the
complex-$N$ plane. Vanishing $\mu_0$ corresponds to the standard minimal
form (\ref{dyres1}). For definiteness, and for simplicity, we choose
$p=2$ in this paper. We will continue to use the expansions
in Eq.~(\ref{lndeltams}), but redefining $\lambda$ in Eq.~(\ref{lamdef})
by
\beq  \label{lamdef1}
\lambda=\b0 \as(\mu) \ln \bar{N}-\frac{1}{2}
\b0 \as(\mu) \ln \left( 1 + \frac{\bar{N}^2\mu_0^2}{M^2} 
\right) \; . 
\eeq
This form has the advantage that it generates only even power corrections
in the ratio $N^2/M^2$, when this quantity is not too large \cite{KSV}. 
We will investigate below the moderation of the perturbative
increase provided by the cut-off $\mu_0$.  

\subsection{Matching to the NLO cross section, and inverse Mellin transform}

As we have discussed above, the resummation is achieved in Mellin
moment space. In order to obtain a resummed cross section in 
$z$ space, one needs an inverse Mellin transform. This
requires a prescription for dealing with the singularity
in the perturbative strong coupling constant in 
Eqs.~(\ref{dyres}),(\ref{dyres1}) or in the 
NLL expansion, Eqs.~(\ref{h1fun}),(\ref{h2fun}). We will use
the Minimal Prescription developed in Ref.~\cite{Catani:1996yz},
which relies on use of the NLL expanded form
Eqs.~(\ref{h1fun}),(\ref{h2fun}),
and on choosing an appropriate contour in the complex-$N$ plane.
In the standard minimal prescription, based on Eqs.\ (\ref{dyres1})
and (\ref{lamdef}), this contour is chosen to lie to the {\it left}
of the ``Landau" singularities at $\lambda=1/2$ in the Mellin integrand,
which are far to the right in the $N$ plane.  With the modified
variable $\lambda$ of Eq.\ (\ref{lamdef1}), however, branch
cuts from $\lambda=1/2$ reside on the imaginary axis 
in the $N$-plane (and so do branch cuts that arise when the argument
of the logarithm in Eq.~(\ref{lamdef1}) vanishes).  
We again choose our inverse contour as 
\begin{align}
\label{hadnmin}
\frac{\tau d\delta\sigma^{\rm (res)}}{d\tau d\phi} &=
\;\int_{C_{MP}-i\infty}^{C_{MP}+i\infty}
\;\frac{dN}{2\pi i} \; \tau^{-N}
\frac{d\delta\sigma^{{\rm (res)},N}}{d\phi} \; ,
\end{align}
where $C_{MP}$ is any positive number.  All singularities
are then  to the left of the contour.
We keep the Mellin contour parallel to the imaginary-$N$ axis. The
result defined by the minimal prescription has the property that 
its perturbative expansion is an asymptotic series that 
has no factorial divergence and therefore
no ``built-in'' power-like ambiguities. Power corrections may
then be added, as phenomenologically required,
and in a sense our cut-off prescription does exactly this.

When performing the resummation, one of course wants to make full
use of the available fixed-order cross section, which in our case
is NLO (${\cal O}(\as)$). Therefore, a matching to this cross 
section is appropriate, which may be achieved by expanding the resummed 
cross section to ${\cal O}(\as)$, subtracting the expanded result
from the resummed one, and adding the full NLO cross section:
\begin{eqnarray}
\label{hadnres}
\frac{\tau d\delta\sigma^{\rm (match)}}{d\tau d\phi}
&=& \sum_{q,\bar{q}}\,\sigma_{q\bar{q}}^{(0)}(\phi)
\;\int_{C_{MP}-i\infty}^{C_{MP}+i\infty}
\;\frac{dN}{2\pi i} \;\tau^{-N} \delta q^N
(\mu^2) \; \delta \bar{q}^N(\mu^2)\nonumber \\
&&\times 
\left[\delta \omega_{q\bar{q}}^{{\rm res},N} (r,\as(\mu))- \left. 
\delta \omega_{q\bar{q}}^{{\rm res},N} (r,\as(\mu)) 
\right|_{{\cal O}(\as)} \, \right] 
\;+\;\frac{\tau d\delta\sigma^{\rm (NLO)}}{d\tau d\phi} \;.
\end{eqnarray}
In this way, NLO is taken into account in full, and the soft-gluon 
contributions beyond NLO are resummed to NLL. Any double-counting
of perturbative orders is avoided. As we will see below, however,
matching is almost academic in the present calculation since
the NLO-expansion of the resummed cross section agrees
to within 0.1\% with the full NLO one for the kinematics
relevant here. We also note that whenever we will use 
the form~(\ref{lamdef1}) with a non-vanishing cut-off $\mu_0$,
we will perform the matching using the expansion
$\left. \delta \omega_{q\bar{q}}^{{\rm res},N} (r,\as(\mu)) 
\right|_{{\cal O}(\as)}$ in~(\ref{hadnres}) evaluated at $\mu_0=0$. 

\section{Phenomenological Results \label{sec4}}

Starting from Eq.~(\ref{hadnres}), we are now ready to present 
some first resummed results at the hadronic level. 
This is not meant to be an exhaustive study; rather
we should like to investigate the overall size and relevance 
of the resummation effects. For this reason, we only
consider the cross section $d^2\sigma/dMd\phi$, 
integrated over all rapidities. This should be sufficient
to study the main effects. In experiment one will eventually
study rapidity distributions in order to better pin down 
the $x$-dependences of the transversity densities. Our resummation
could be extended to this case using techniques developed in~\cite{thrrap}. 
We also note that the charmonium 
resonances will dominate over a part of the spectrum that we
will consider. For our case study we will just ignore this,
but remind the reader that a complete treatment of the dilepton
spectrum will eventually also require the incorporation of charmonium 
production and its resummation effects.

We will investigate $\bar{p}p$ collisions at four different energies.
Each of these may eventually be realized at the GSI. The first two are in 
the fixed-target mode, relevant for the initial stage of $\bar{p}p$
physics at the GSI, when antiprotons will be scattered off
proton targets at energies in the range $15$~GeV $\lesssim E_{\bar{p}}
\lesssim 25$~GeV. For definiteness, we will consider 
$S=30$~GeV$^2$ and $S=45$~GeV$^2$.
This is also the regime in which the theoretical description 
is most challenging. The second regime is for an asymmetric collider,
with $E_p=3.5$~GeV and $E_{\bar{p}}=15$~GeV, corresponding to 
$\sqrt{S}=14.5$~GeV. Finally, we will consider a symmetric collider
with $E_p=E_{\bar{p}}=15$~GeV ($\sqrt{S}=30$~GeV). In all 
calculations below, we choose the renormalization and factorization 
scales as $\mu=M$.

\subsection{Unpolarized cross section near partonic threshold}

We start by considering the unpolarized cross section. For
some of the kinematic regions described above, Drell-Yan experiments
at the GSI would enter uncharted territory, and it will
be crucial to develop confidence that the theoretical framework
is understood. An important test would be a comparison to
precise measurements of the unpolarized cross section. 

For our unpolarized calculations we will use the NLO 
($\msbar$ scheme) GRV parton distributions throughout~\cite{grv}. 
In the unpolarized case, we are in the fortunate 
situation that even the full NNLO corrections to the partonic Drell-Yan 
cross section are available~\cite{vN}, which we will incorporate in
our studies. These should in principle be used in 
conjunction with a set of NNLO parton distributions, which
became available recently~\cite{mrstnnlo} after the 
computation of the three-loop evolution kernels~\cite{3loop}.
The main purpose of our present studies, however, is to see
how well the soft-gluon terms we are resumming reproduce also the 
NNLO corrections to the cross section. For this it is sufficient  
to stick to our use of NLO parton densities.  

\begin{figure}[t!]
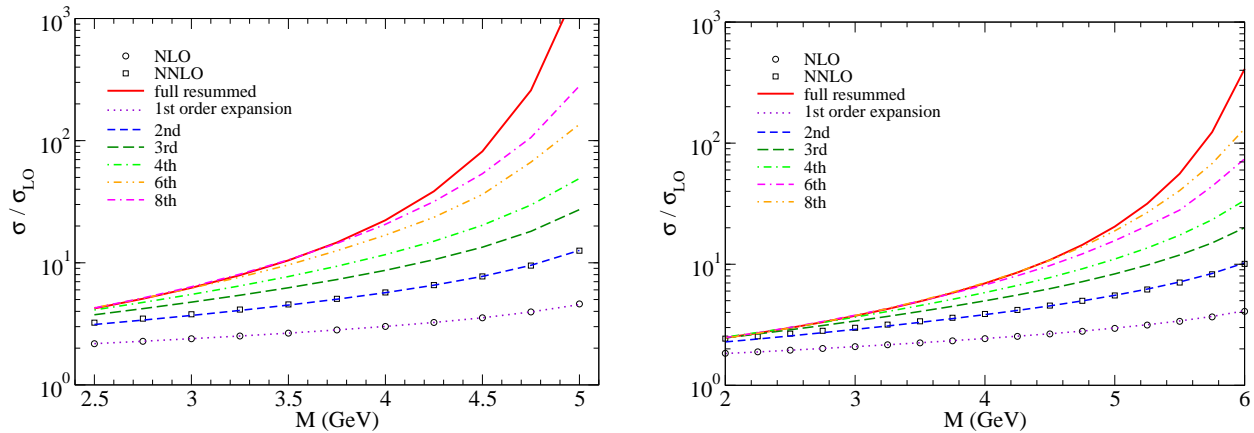

\begin{center}
\vspace*{0.8cm}
\hspace*{-5mm}
\epsfig{figure=kfac-fixed.eps,width=0.45\textwidth}
\hspace*{5mm}
\epsfig{figure=kfac-s45.eps,width=0.45\textwidth}
\end{center}
\vspace*{-.5cm}
\caption{``$K$-factors'' relative to LO as defined in 
Eqs.~(\ref{eq:kres}) and~(\ref{ksoftg}) for the Drell-Yan 
cross section in fixed-target $\bar{p}p$ collisions at
$S=30$~GeV$^2$ (left) and $S=45$~GeV$^2$ (right), as functions of
lepton pair invariant mass $M$. The symbols denote
the results for the exact NLO and NNLO calculations. \label{fig1}}
\vspace*{0.cm}
\end{figure}

Figure~\ref{fig1} shows the effects of the higher orders
generated by resummation, for the fixed-target cases with
$S=30$~GeV$^2$ and $S=45$~GeV$^2$. We define a resummed ``$K$-factor'' as 
the ratio of the resummed (matched) cross section to the LO cross section,
\begin{equation}
\label{eq:kres}
K^{{\rm (res)}} = \f{{d\sigma^{\rm (match)}}/{dMd\phi}}
{{d\sigma^{\rm (LO)}}/{dMd\phi}}\, ,
\end{equation}
which is shown by the solid line in Fig.~\ref{fig1}. 
As can be seen, $K^{{\rm (res)}}$ is very large, meaning that resummation
results in a dramatic enhancement over LO,
sometimes by over two orders of magnitude.
It is then interesting to see how this enhancement builds up
order by order in the resummed cross section. We 
expand the matched resummed formula to NLO and beyond and define the 
(here, not matched) ``soft-gluon $K$-factors''
\begin{equation} \label{ksoftg}
K^n\;\equiv\; \f{{\left. d\sigma^{\rm{(res)}}/{dMd\phi}\right|_{{\cal O}
(\as^{n})}}}{{d\sigma^{\rm (LO)}}/{dMd\phi}} \; ,
\end{equation}
which for $n=1,2,\ldots$ give the additional enhancement due to 
the ${\cal O}(\as^{n})$ terms in the resummed formula. Formally,
$K^0=1$, while $K^{\infty}=K^{{\rm (res)}}$ up to the effects
of matching at NLO. The 
results for $K^{1,2,3,4,6,8}$ are also shown in Fig.~\ref{fig1}. 
One can see that there are very large contributions even 
beyond NNLO, in particular at the higher $M$. Clearly, the
full resummation given by the solid line
receives contributions from high orders.
\begin{figure}[t!]
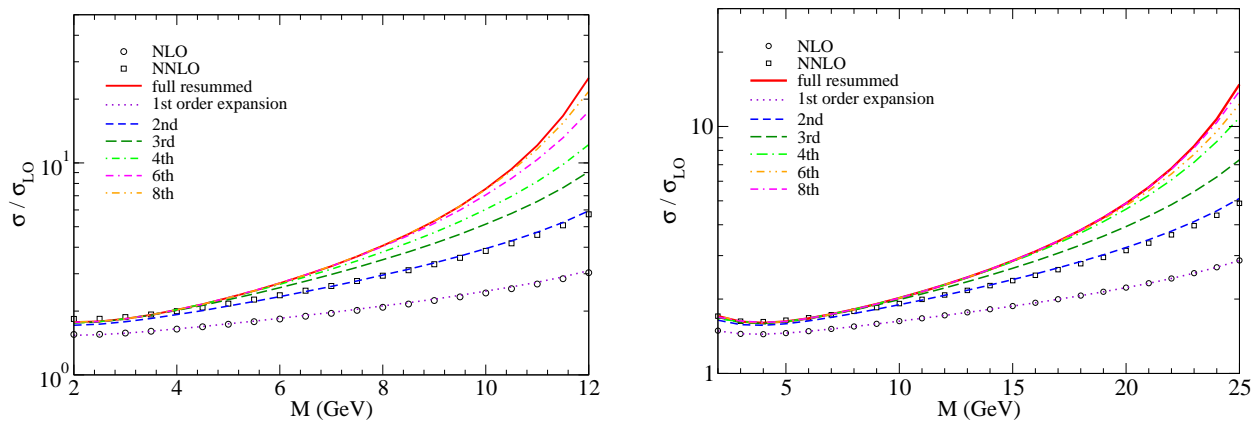

\begin{center}
\vspace*{0.8cm}
\hspace*{-5mm}
\epsfig{figure=kfac-s210.eps,width=0.45\textwidth}
\hspace*{5mm}
\epsfig{figure=kfac-coll.eps,width=0.45\textwidth}
\end{center}
\vspace*{-.5cm}
\caption{Same as Fig.~\ref{fig1}, but for $\bar{p}p$
collider options with $\sqrt{S}=14.5$~GeV (left) and 
$\sqrt{S}=30$~GeV (right).  \label{fig2}}
\vspace*{0.cm}
\end{figure}
The symbols in Fig.~\ref{fig1} show the associated $K$ factors 
for the exact NLO and NNLO calculations, respectively. One can see
that these agree extremely well with the ${\cal O}(\alpha_s)$ and 
${\cal O}(\alpha_s^2)$ expansions of the resummed cross section. In fact, 
the agreement between the full NLO result and the ${\cal O}(\alpha_s)$ 
expansion of the resummed cross section is better than 0.1\% over
the whole range in $M$ shown. 
Thus the matched resummed $K$-factor 
$K^{{\rm (res)}}$ of Eq.~(\ref{eq:kres})  is
also numerically very close to $K^{\infty}$ as
defined in~(\ref{ksoftg}).
We note that the replacement in
Eq.~(\ref{evol1}) helps somewhat in this comparison, in particular
at NNLO  it leads to a relative improvement of a few per cent. 
From the comparison we may conclude that the terms that we resum to all 
orders strongly dominate the cross section.

In Fig.~\ref{fig2}, we show similar results for the two collider
modes at $\sqrt{S}=14.5$ and 30~GeV. One can see that for a 
fixed Drell-Yan mass $M$ the corrections become much smaller, since 
one is much further away from partonic threshold than for the 
fixed-target cases. Also, the convergence of the perturbative
series occurs somewhat more rapidly. As before, the agreement
between the ${\cal O}(\alpha_s)$ and ${\cal O}(\alpha_s^2)$ 
expansions of the resummed cross section and the exact NLO and NNLO 
calculations is very good, demonstrating the relevance of the resummed result.

\begin{figure}[t!]
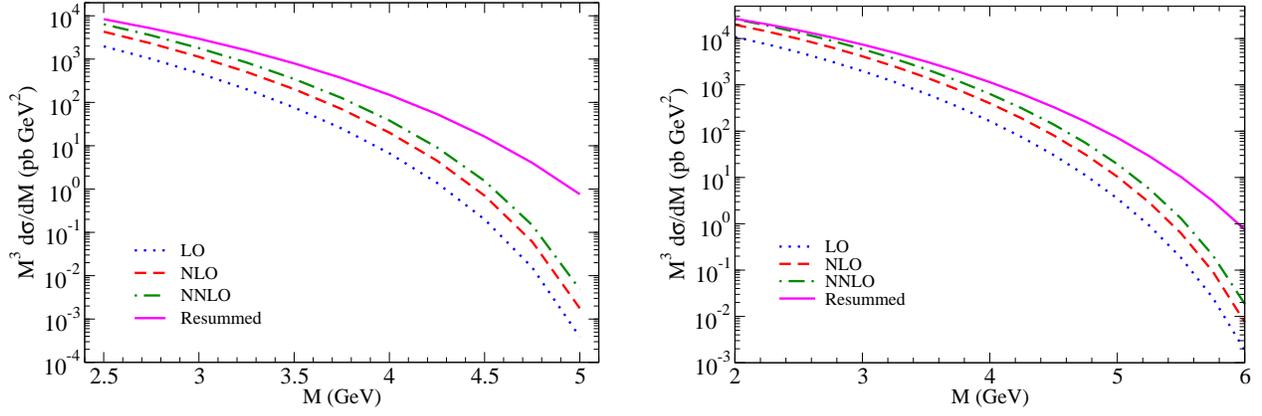

\begin{center}
\vspace*{0.8cm}
\hspace*{-5mm}
\epsfig{figure=fixed-unp.eps,width=0.45\textwidth}
\hspace*{5mm}
\epsfig{figure=s45_unp.eps,width=0.45\textwidth}
\end{center}
\vspace*{-.5cm}
\caption{Unpolarized cross sections $d^2\sigma/dM$ 
at $S=30$~GeV$^2$ (left) and $S=45$~GeV$^2$ (right) at LO, NLO, 
NNLO, and NLL resummed, as functions of
lepton pair invariant mass $M$. \label{fig3}}
\vspace*{0.cm}
\end{figure}

\begin{figure}[t!]
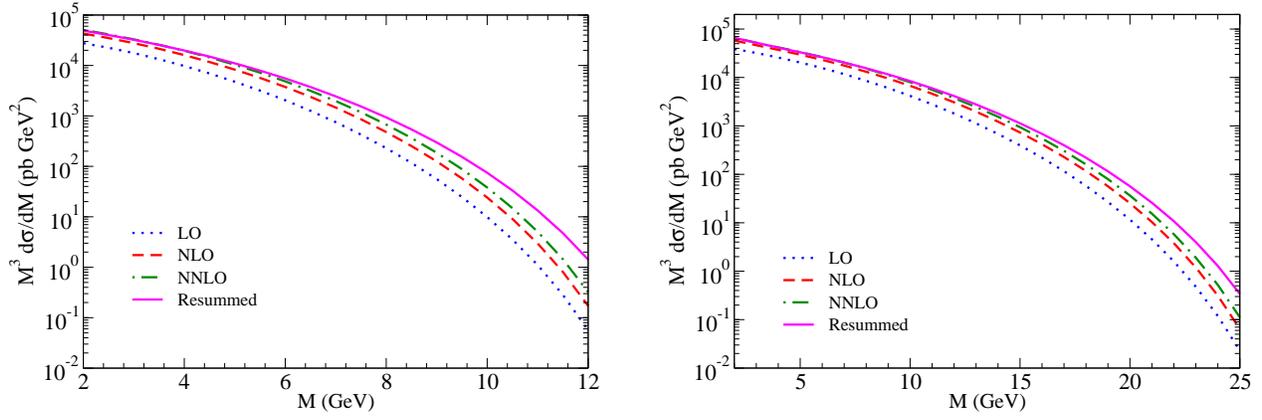

\begin{center}
\vspace*{0.8cm}
\hspace*{-5mm}
\epsfig{figure=s210-unp.eps,width=0.45\textwidth}
\hspace*{5mm}
\epsfig{figure=coll-unp.eps,width=0.45\textwidth}
\end{center}
\vspace*{-.5cm}
\caption{Same as Fig.~\ref{fig3}, but for $\bar{p}p$
collider options with $\sqrt{S}=14.5$~GeV (left) and 
$\sqrt{S}=30$~GeV (right). \label{fig4}}
\vspace*{0.cm}
\end{figure}

In Figs.~\ref{fig3} and~\ref{fig4} we show the actual unpolarized cross 
sections $M^3 d\sigma/dM$ corresponding to the results
shown in Figs.~\ref{fig1} and~\ref{fig2}, at (exact) fixed orders
(LO, NLO, NNLO) in perturbation theory, and for the NLL resummed
case. Here we have integrated over all azimuthal angles 
$\phi$\footnote{Note that the normalization factor $\sigma_{q\bar{q}}^{(0)}$
in Eq.~(\ref{eq:lo}) becomes $2\alpha^2e_q^2/9S$ in the 
unpolarized case.}. 


\begin{figure}[t!]
\begin{center}
\vspace*{1cm}
\epsfig{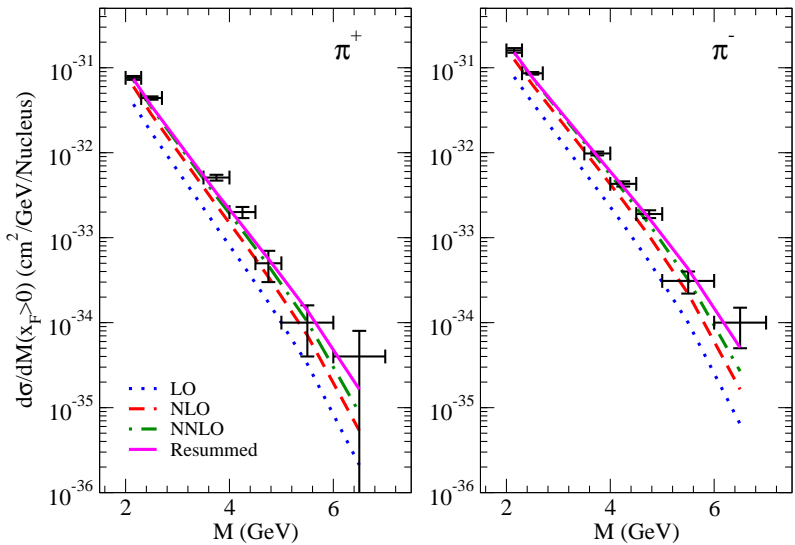}
\end{center}
\vspace*{-.5cm}
\caption{Comparison of LO, NLO, NNLO and resummed results to the
WA39 data~\cite{wa39} for Drell-Yan dimuon production in 
$\pi^{\pm}W$ scattering. \label{fig5}}
\vspace*{0.cm}
\end{figure}

One may wonder whether any sign of the need for large corrections beyond
NLO can be found in previous Drell-Yan data~\cite{dyrev}. 
The measurements at the
lowest $\sqrt{S}$ we are aware of are from the CERN WA39~\cite{wa39} 
experiment and were made in $\pi^{\pm}$-Tungsten scattering. The pion 
energy was $E_{\pi}=39.5$~GeV, higher than what is considered for the 
fixed-target mode at the GSI, but still quite far below the collider 
energies. Fig.~\ref{fig5} shows the data, along with our results at 
LO, NLO, NNLO and NLL-resummed. We are using the parton distributions 
for the pion of~\cite{grvpi}. It is hard to draw definite conclusions 
from the comparison in  Fig.~\ref{fig5}, partly because the experimental 
uncertainties are rather large, and also because the pion parton densities
are not known accurately. Nevertheless, it is interesting to observe that
the data points at the highest dimuon mass $M$ are quite consistent with 
large perturbative resummation effects. 

\begin{figure}[t!]
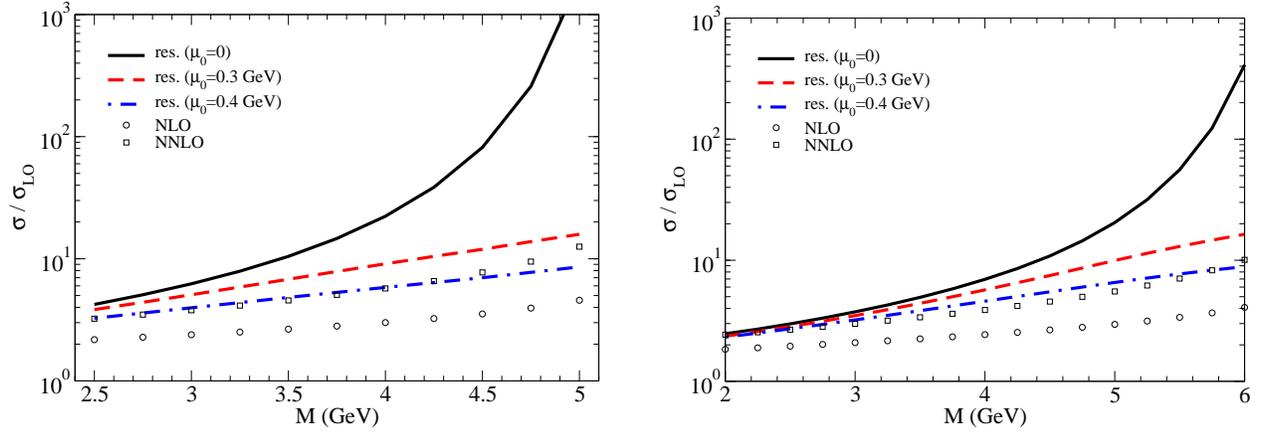

\begin{center}
\vspace*{0.8cm}
\hspace*{-5mm}
\epsfig{figure=kfac-ps-s30-rev.eps,width=0.45\textwidth}
\hspace*{5mm}
\epsfig{figure=kfac-ps-s45-rev.eps,width=0.45\textwidth}
\end{center}
\vspace*{-.5cm}
\caption{``$K$-factors'' relative to LO as in Fig.~\ref{fig1}, at
$S=30$~GeV$^2$ (left) and $S=45$~GeV$^2$ (right). The dashed lines
show the effect of a lower cut-off $\mu_0=300$~MeV for the $k_T$-integral 
in the exponent, and the dot-dashed a cut-off
of $\mu_0=400$~MeV for comparison.  The symbols denote
the results for the exact NLO and NNLO calculations. \label{fig1a}}
\vspace*{0.cm}
\end{figure}

\begin{figure}[t!]
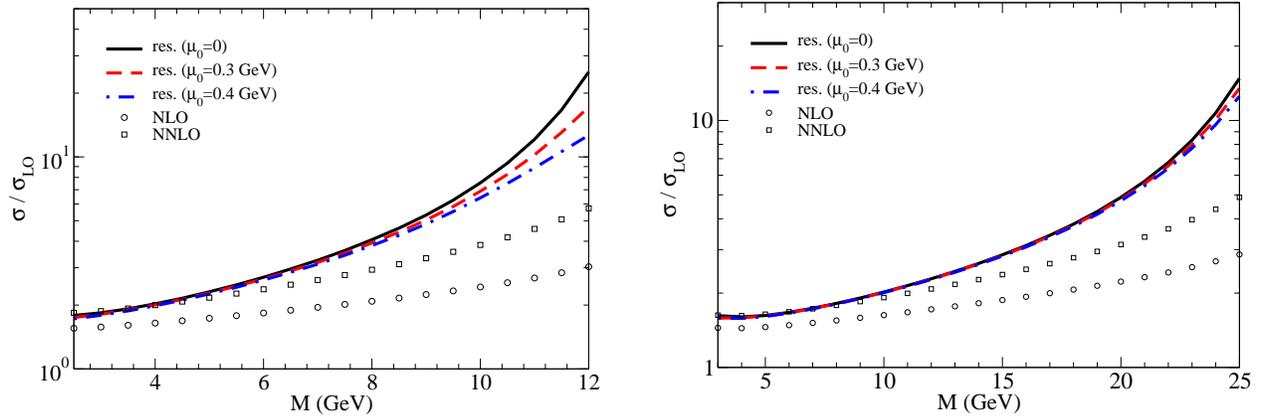

\begin{center}
\vspace*{0.8cm}
\hspace*{-5mm}
\epsfig{figure=kfac-ps-s210-rev.eps,width=0.45\textwidth}
\hspace*{5mm}
\epsfig{figure=kfac-ps-s900-rev.eps,width=0.45\textwidth}
\end{center}
\vspace*{-.5cm}
\caption{Same as Fig.~\ref{fig1a}, but for $\bar{p}p$
collider options with $\sqrt{S}=14.5$~GeV (left) and 
$\sqrt{S}=30$~GeV (right).  \label{fig2a}}
\vspace*{0.cm}
\end{figure}

Given the large, not to say huge, size of some of the enhancements,
we now turn to the same ratios computed using the
exponent~(\ref{dyres2}) with the modified lower limit
on the $k_T$ integral,
which regulates its infrared behavior.  For simplicity, we
will only show results for relatively small values of the 
cut-off scale $\mu_0$ in (\ref{lamdef1}).
Of course, different choices give different results,
but we should think of $\mu_0$ as a kind of factorization scale, separating
perturbative contributions from nonperturbative.  Thus changes in
$\mu_0$ would be compensated 
at least in part by changes in a nonperturbative function.
On the other hand, a very strong sensitivity to $\mu_0$ 
can reasonably be interpreted as indicating that 
perturbative resummation alone cannot give
a reliable estimate for the cross section.
Our interest here, therefore, is primarily to illustrate the modification of
the perturbative sector, which we do by choosing $\mu_0=0.3$ GeV
and $0.4$ GeV.   
Results for the ``$K$-factor" with these values of $\mu_0$ are shown
in Figs.\ \ref{fig1a} and \ref{fig2a}, compared to the same NLO, NNLO and
resummed factors shown above at the relevant energies
and pair masses $M$.   We hope to study the $\mu_0$ dependence
more extensively elsewhere, in connection with 
possible nonperturbative corrections.

The ratios of the new resummed but infrared-regulated 
cross sections to the LO one show a smoother increase
than the pure minimal resummed cross sections.  This difference is
particularly marked at the lower center-of-mass energies
in Fig.\ \ref{fig1a}, with only a modest enhancement over NNLO
remaining at $\mu_0=0.3$ GeV, and even lower at $0.4$ GeV. 
We note that the NLO and NNLO expansions of the
resummed cross section turn out to be much less affected
by the cut-off $\mu_0$ than the full resummed cross section. 
At the higher energies of  Fig.\ \ref{fig2a}, the
regulated resummed curves follow the unregulated curves
far above NNLO, with much reduced sensitivity to $\mu_0$.  
We interpret these results to indicate a
strong sensitivity to nonperturbative dynamics at the lower energies,
and much less at the higher.   
We therefore take the predictions
of large enhancements due to high order perturbation theory
more seriously in the latter case, even while keeping in mind
the WA39 measurements of Fig.~\ref{fig5} above, 
which suggest that large corrections,
perturbative or not, should not be ruled out a priori.  
Data over the entire  kinematic regime considered here would
certainly shed a unique light on the transition between 
long- and short-distance effects in hadronic scattering.

\subsection{Spin asymmetry $A_{TT}$}

Before we can perform numerical studies of $A_{TT}$ we need to make 
a model for the transversity densities in the valence region. Here, 
guidance is provided by the Soffer inequality \cite{ref:soffer}
\begin{equation}
\label{eq:sofferineq}
2\left|\delta q(x,Q^2)\right| \leq q(x,Q^2) + \Delta q(x,Q^2) \; ,
\end{equation}
which gives an upper bound for each $\delta q$. 
Following~\cite{ref:attlo,ref:attnlo} we utilize this inequality by saturating 
the bound at some low input scale $Q_0\simeq 0.6\,\mathrm{GeV}$ using 
the NLO GRV \cite{grv} and GRSV (``standard scenario'') 
\cite{grsv} densities $q(x,Q_0^2)$ and $\Delta q(x,Q_0^2)$,
respectively. For $Q>Q_0$ the transversity densities $\delta q(x,Q^2)$ 
are then obtained by solving the evolution equations with the
NLO~\cite{ref:dy2,ref:nlokernels} evolution kernels. We refer
the reader to \cite{ref:attlo,ref:attnlo} for more details on our 
model distributions.

\begin{figure}[t!]
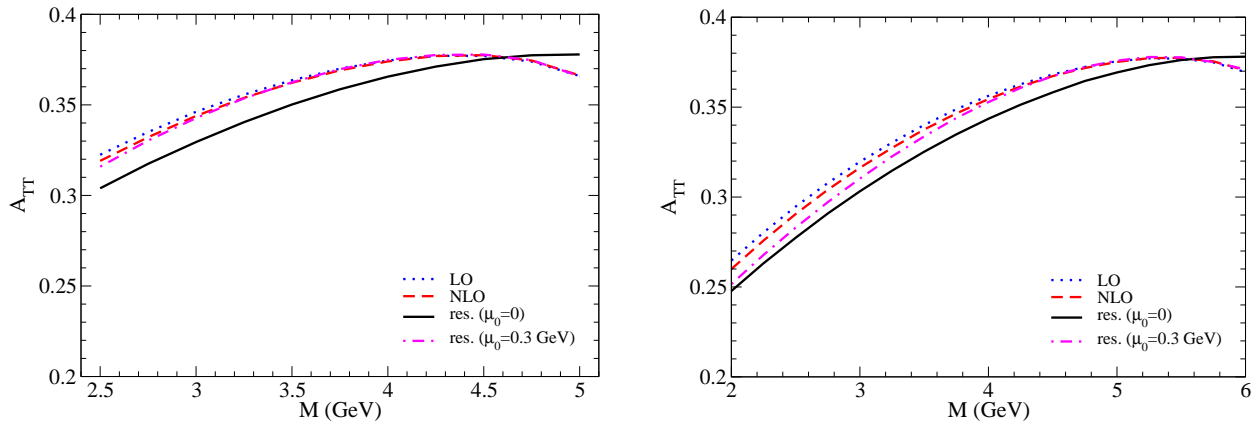

\begin{center}
\vspace*{0.8cm}
\hspace*{-5mm}
\epsfig{figure=asym-s30.eps,width=0.45\textwidth}
\hspace*{5mm}
\epsfig{figure=asym-s45.eps,width=0.45\textwidth}
\end{center}
\vspace*{-.5cm}
\caption{Spin asymmetry $A_{TT}(\phi=0)$ at LO, NLO and for the NLL
resummed case, at $S=30$~GeV$^2$ (left) and $S=45$~GeV$^2$ (right). 
The dash-dotted lines show the effect of a lower
cut-off $\mu_0=300$~MeV for the $k_T$-integral in the
exponent. \label{fig6}}
\vspace*{0.cm}
\end{figure}

\begin{figure}[t!]
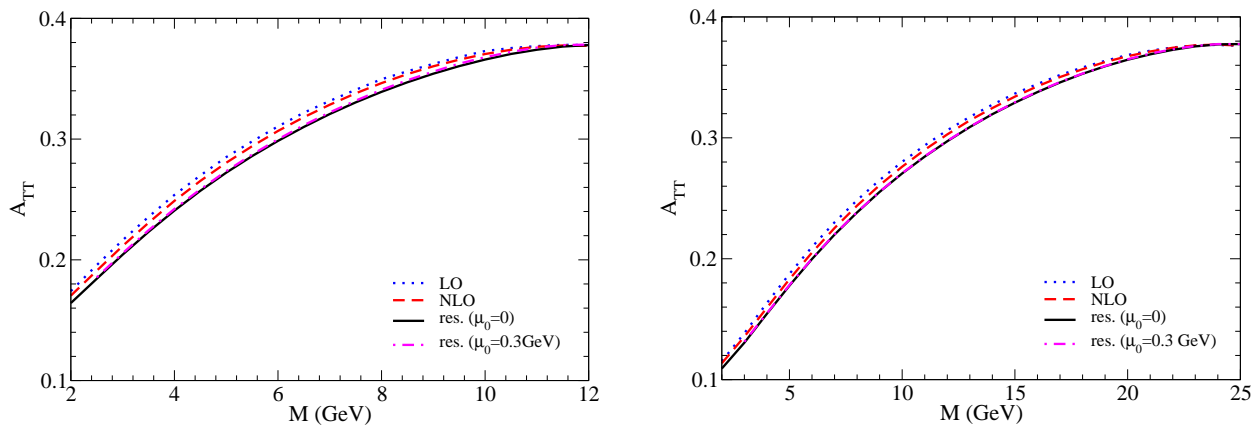

\begin{center}
\vspace*{0.8cm}
\hspace*{-5mm}
\epsfig{figure=asym-s210.eps,width=0.45\textwidth}
\hspace*{5mm}
\epsfig{figure=asym-s900.eps,width=0.45\textwidth}
\end{center}
\vspace*{-.5cm}
\caption{Same as Fig.~\ref{fig6}, but for $\bar{p}p$
collider options with $\sqrt{S}=14.5$~GeV (left) and 
$\sqrt{S}=30$~GeV (right). \label{fig7}}
\vspace*{0.cm}
\end{figure}

We will now investigate to what extent the large perturbative
corrections we found for the unpolarized Drell-Yan cross section 
cancel in the spin asymmetry $A_{TT}$, Eq.\ (\ref{attdefi}). 
We have mentioned earlier that the
resummation factors~(\ref{dyres}) for the $q\bar{q}$ cross section 
are the same in the unpolarized and transversely polarized
cases if both are treated in the same factorization scheme. 
Resummation effects would therefore cancel if the spin 
asymmetry were in Mellin-moment space. The convolution with the
parton distributions and the inverse Mellin transform will
affect the cancellation somewhat, but one still expects 
the spin asymmetry to be very robust. Indeed, the results
for $A_{TT}$ at LO, NLO, and resummed to NLL (with and
without the cut-off $\mu_0$), shown in Figs.~\ref{fig6} 
and~\ref{fig7} for the four energies that we consider, confirm this. 
We show 
\begin{equation} \label{eqatt}
A_{TT}=\frac{d\delta \sigma/dM d\phi}{d\sigma/dMd\phi}
\end{equation}
as a function of $M$. For simplicity we set $\phi=0$ here; extension 
to other $\phi$ is straightforward by taking into account the 
$\cos(2\phi)$-dependence displayed in Eq.~(\ref{eq:lo}).
 
Aside from a slight deficit in the curves at lower $M$
(and slight excess at higher $M$) with unmodified 
minimal resummation at the lower  $S$, all curves, including the LO, NLO and 
regulated resummed asymmetries all lie within a few percent
of each other. We note that to NLO this robustness of $A_{TT}$ was 
also found for $\bar{p}p$ collisions at higher energies~\cite{ratcl04}.


\begin{figure}[htb]
\begin{center}
\vspace*{0.9cm}
\epsfig{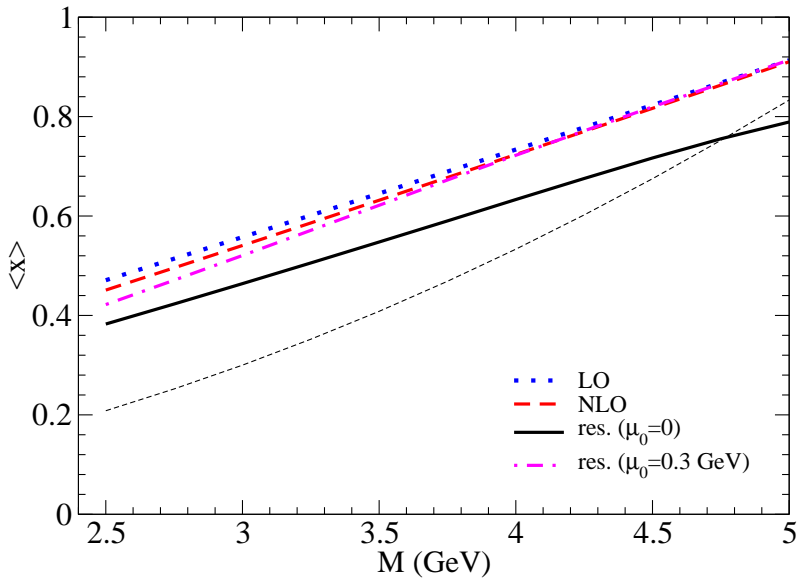}
\end{center}
\vspace*{-.5cm}
\caption{Average values $\langle x_a\rangle=\langle x_b\rangle\equiv
\langle x\rangle$ for the parton momentum fractions at
$S=30$~GeV$^2$ as functions of $M$, at LO, NLO, for purely perturbative
resummed, and for regulated resummed with $\mu_0=300$~MeV. For
comparison, the lowest curve shows the corresponding
values of $\tau=M^2/S$. A curve for $\sqrt{\tau}$ -- a straight line
in this plot -- would be 
practically indistinguishable from the LO one, although remaining
slightly below it. \label{fig:xav}}
\vspace*{0.cm}
\end{figure}

We can shed some light on why the asymmetry
is modestly but significantly shifted for the unregulated
perturbative resummed cross section with respect to
the other cases shown.
First, we observe that smaller momentum fractions $x_{a,b}$ in the
factorized  cross section are associated with
the cross section at smaller pair mass $M$,
where the spin asymmetry is slightly smaller.
Therefore, any effect that tends to  drive momentum fractions
particularly close to their minimum values, at $z=1$,  will tend
to decrease the asymmetry.
We note, however, that because the valence quark distributions are
decreasing functions of the  $x$'s, and because they are
the same for both hadrons, we might anticipate that the average values 
$\langle x_a \rangle=\langle x_b \rangle$
are not far from  their symmetric values,  $x_a=x_b=\sqrt{\tau}$ at  
partonic threshold.
In fact, this turns out to be a surprisingly accurate estimate, even at 
lowest order, as can be readily verified from Fig.~\ref{fig:xav}, 
where we plot $\langle x\rangle$
for LO, NLO, and the unregulated ($\mu_0=0$) and regulated
($\mu_0=300$~MeV) resummed cross sections.
In this figure $\langle x\rangle$ is found by
performing the integral for the unpolarized cross section
with an extra factor $x_a$ in the integrand,
and by dividing by the cross section itself.

Beyond lowest order, the hard-scattering cross section
is further enhanced at partonic threshold ($z=1$), and we would expect 
that large
perturbative enhancements, such as those associated with the plus 
distribution of Eq.\ (\ref{cq1z}),
would force the average partonic center-of-mass energy
still closer to threshold, and 
hence reduce the asymmetry even further.
We should observe, however, that the hard scattering function is
not a positive-definite cross section, but rather a sum of plus 
distributions,
given at first nontrivial order in Eq.\ (\ref{cq1z}).   At NLO, for 
example,
the positive contribution at $z=1$ is from a delta function associated
with virtual corrections, while the real-gluon contribution, 
$\ln(1-z)/(1-z)$,
is actually negative, due to the subtraction of
collinear divergences in the calculation of the hard scattering 
\cite{dyresum1,dyresum2}.
We therefore cannot interpret $\langle x_{a,b}\rangle$ as averages in the
usual sense.  In any case, we  do see a more significant decrease in
$\langle x \rangle$, computed above, for the unregulated resummed cross 
section than
for fixed order.  In fact, the values derived in this manner are below 
$\sqrt{\tau}$,
which would be  the lower limit for a positive-definite hard scattering 
function, and at the highest $M$, even below $\tau$, which
is the lower limit of the integration range for the  $x$'s.
The caveat against a literal interpretation of $\langle x\rangle$ 
notwithstanding,
it is reasonable to interpret the
modest decrease in the asymmetry for the unregulated resummed
cross section as resulting from a hard-scattering function that is
exceptionally peaked near $z=1$ in this case.

\subsection{Use of DIS parton distribution functions}

\begin{figure}[htb]
\begin{center}
\vspace*{0.8cm}
\epsfig{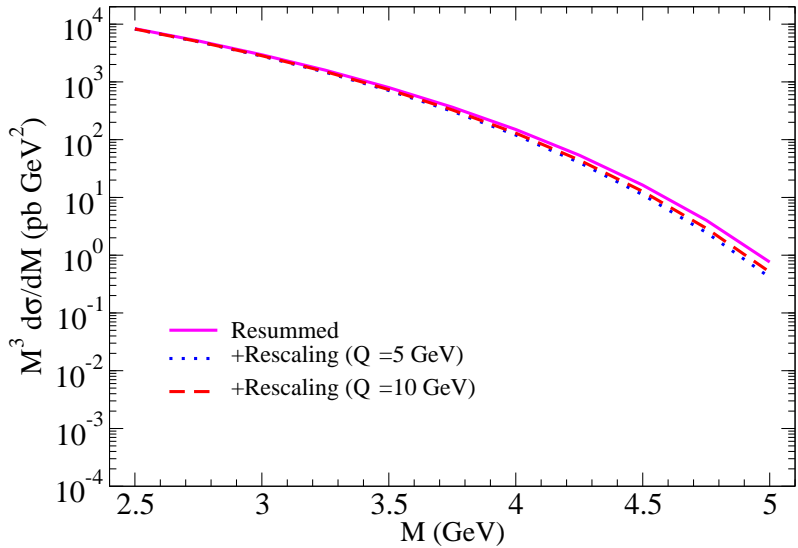}
\end{center}
\vspace*{-.5cm}
\caption{Effect on the unpolarized Drell-Yan cross
section at $S=30$~GeV$^2$ due to ``rescaling'' of the parton 
distributions as in Eq.~(\ref{fresdef}). The solid line shows
the result for the resummed cross section as in Fig.~\ref{fig3}, 
the dotted and dashed lines are for $Q=5$ and 10~GeV, respectively,
in Eq.~(\ref{fresdef}). \label{fig8}}
\vspace*{0.cm}
\end{figure}

We recall that throughout this study, we have used 
the NLO parton distribution functions of Ref.\ \cite{grv}.
Especially given the large effects we have found with
resummation, we should revisit the use of ``un-resummed''
parton distributions in our study. 
The momentum fractions 
probed in the parton distributions become very large in 
the fixed-target regime, as shown by Fig.~\ref{fig:xav}. 
One may wonder here to what extent
the parton distributions themselves should include resummation
effects. The densities we use have been determined mostly from an
analysis of data from deeply-inelastic scattering (DIS), in which 
however no resummation of large-$N$ logarithms was included.
It is known that soft-gluon resummation effects in DIS are rather
unimportant, except at moderate photon virtuality 
$Q^2$ and very large $x$~\cite{stwvpdf,largexdis}.  
Nonetheless, we will give a rough estimate of the quantitative 
effect on the Drell-Yan cross section that might occur {\it if} one 
used parton densities determined from an analysis of the DIS data 
including resummation. We follow~\cite{stwvpdf} to determine a model 
set of ``$\msbar$-resummed'' valence distributions $q^{N,{\rm res}}$ (in 
Mellin-moment space) by demanding that their contributions to the 
DIS structure function $F_2$ match those of the corresponding NLO 
densities at a fixed scale $Q$. This is ensured by ``rescaling''
the parton densities:
\beq
q^{N,{\rm res}}(Q^2)=q^{N,{\rm NLO}}(Q^2)\;
{C_2^{\rm NLO}(N,Q^2) \over 
C_2^{\rm res}(N,Q^2)}\, ,
\label{fresdef}
\eeq
where $C_2^{\rm NLO}$ and $C_2^{\rm res}$ are the perturbative NLO
and NLL resummed quark coefficient functions for $F_2$, respectively, 
which may be found in~\cite{stwvpdf} for example. We choose $Q$
in Eq.~(\ref{fresdef}) fairly large, so that it is in a region 
where resummation effects are expected to be small and NLO to yield
a good description of $F_2$. For illustration, we use two different
values, $Q^2=25$~GeV$^2$ and $Q^2=100$~GeV$^2$. The 
ansatz~(\ref{fresdef}) then represents an estimate of the likely change 
in the parton distributions from resummation in DIS, and the
parton densities $q^{N,{\rm res}}$ may be used for calculating
the Drell-Yan cross section. The result is shown in Fig.~\ref{fig8},
where we repeat the resummed cross section from Fig.~\ref{fig3},
and display the effects on the cross section for the two
choices of $Q$. A moderate decrease of the resummed cross section
is found. We note that this effect becomes smaller at the collider 
energies, for a given $M$.

\section{Conclusions}

We have verified that perturbative corrections associated
with partonic threshold are large for
dilepton production in the kinematic region being considered
for proton-antiproton collisions at GSI.  The close agreement
between the fixed-order expansions of threshold-resummed
cross sections and exact fixed order calculations suggests that the
resummation is physically relevant, and significantly enhances the
unpolarized Drell-Yan cross sections beyond fixed orders. At the same 
time, the transverse-spin asymmetries whose measurement is suggested
for these experiments, are remarkably insensitive to
shifts in the overall normalization.  In summary, perturbative
corrections appear to make the cross sections larger
independently of spin.  They would
therefore make easier the study of spin asymmetries,
and ultimately transversity distributions.

We have also shown that at the lower energies considered, the resummed
cross section decreases markedly when an infrared cut-off is used to
regulate contributions from soft gluon emission in
the far infrared region.  We have shown how to 
incorporate such a cut-off as a generalization of
minimal resummation.
At the lower energies, a strong sensitivity to the cut-off
suggests that the large enhancements found with
unregulated perturbative resummation arise
from an unwarranted
extension of perturbation theory into the
soft region  
At the same time, it is interesting to note that
quite substantial $K$-factors survive almost unchanged by infrared regulation
at the higher energies considered. 
This suggests that
the measurement of the unpolarized dilepton cross section
will shed light on the relationship between fixed orders,
perturbative resummation and nonperturbative dynamics
in hadronic scattering.

\section*{Acknowledgments}

We thank D.\ de\ Florian, W.\ Kilgore, M.\ Stratmann
for helpful discussions, and N.~Nikolaev for 
encouragement.
H.S. and H.Y. would like to thank J.~Kodaira
for useful comments and encouragement.
H.Y. is also grateful to the RIKEN BNL Research Center
for helpful hospitality during his stay.
W.V.\ is grateful to RIKEN, Brookhaven National Laboratory
and the U.S.\ Department of Energy (contract number DE-AC02-98CH10886) for
providing the facilities essential for the completion of his work.
The work of G.S.\ was supported in part
by the National Science Foundation, grants PHY-0098527, PHY-0354776,
and PHY-0354822.



\begin{thebibliography}{99}

\bibitem{ref:jaffeji} R.L.\ Jaffe and X.\ Ji, 
Phys. Rev. Lett. {\bf 67}, 552 (1991);
Nucl. Phys. B {\bf 375}, 527 (1992).

\bibitem{hermest}  A.~Airapetian {\it et al.}  [HERMES Collaboration],
Phys.\ Rev.\ Lett.\  {\bf 94}, 012002 (2005) [arXiv:hep-ex/0408013].

\bibitem{ref:ralston} J.P.\ Ralston and D.E.\ Soper,
Nucl. Phys. B {\bf 152}, 109 (1979).

\bibitem{ref:artru} X.\ Artru and M.\ Mekhfi, Z. Phys. C {\bf 45}, 669 (1990).

\bibitem{ref:ratcliffe} A comprehensive review on transversity can be 
found in:  V.~Barone, A.~Drago and P.~G.~Ratcliffe,
Phys.\ Rept.\  {\bf 359}, 1 (2002)
[arXiv:hep-ph/0104283].

\bibitem{ref:fact1} 
J.C.\ Collins, D.E.\ Soper, and G.\ Sterman, 
Nucl.\ Phys.\ B {\bf 261}, 104 (1985);
Nucl.\ Phys.\ B {\bf 308}, 833 (1988); 
in: ``Perturbative Quantum Chromodynamics'', A.H.~Mueller (ed.), 
World Scientific Publ., Singapore, 1989, p.1 [arXiv:hep-ph/0409313];
G.~T.~Bodwin, Phys.\ Rev.\ D {\bf 31}, 2616 (1985)
[Erratum-ibid.\ D {\bf 34}, 3932 (1986)].

\bibitem{ref:fact2} J.C.~Collins, Nucl.\ Phys.\ B {\bf 394}, 
169 (1993) [arXiv:hep-ph/9207265].

\bibitem{ref:dylo} 
J.L.\ Cortes, B.\ Pire, and J.P.\ Ralston,
Z. Phys. C {\bf 55}, 409 (1992);
C.~Bourrely and J.~Soffer, Nucl.\ Phys.\ B {\bf 445}, 341 (1995)
[arXiv:hep-ph/9502261];
V.~Barone, T.~Calarco and A.~Drago, Phys.\ Rev.\ D {\bf 56}, 527 (1997)
[arXiv:hep-ph/9702239];
O.~Martin and A.~Sch\"{a}fer, Z.\ Phys.\ A {\bf 358}, 429 (1997) 
[arXiv:hep-ph/9607470]; 
M.~Miyama, Nucl.\ Phys.\ Proc.\ Suppl.\  {\bf 79}, 620 (1999)
[arXiv:hep-ph/9905559].

\bibitem{ref:dy1}
O.~Martin, A.~Sch\"{a}fer, M.~Stratmann and W.~Vogelsang,
Phys.\ Rev.\ D {\bf 57}, 3084 (1998) [arXiv:hep-ph/9710300].

\bibitem{ref:dy1a} 
O.~Martin, A.~Sch\"{a}fer, M.~Stratmann and W.~Vogelsang,
Phys.\ Rev.\ D {\bf 60}, 117502 (1999) [arXiv:hep-ph/9902250].

\bibitem{ref:attold} K.\ Hidaka, E.\ Monsay, and D.\ Sivers,
Phys. Rev. D {\bf 19}, 1503 (1979);
X.\ Ji, Phys. Lett. B {\bf 284}, 137 (1992).

\bibitem{ref:jaffesaito} R.~L.~Jaffe and N.~Saito,
Phys.\ Lett.\ B {\bf 382}, 165 (1996) [arXiv:hep-ph/9604220].

\bibitem{ref:attlo} J.~Soffer, M.~Stratmann and W.~Vogelsang,
Phys.\ Rev.\ D {\bf 65}, 114024 (2002) [arXiv:hep-ph/0204058].

\bibitem{ref:attnlo} A.~Mukherjee, M.~Stratmann and W.~Vogelsang,
Phys.\ Rev.\ D {\bf 67}, 114006 (2003)  [arXiv:hep-ph/0303226].

\bibitem{review} See, for example:  G.~Bunce, N.~Saito, J.~Soffer 
and W.~Vogelsang, Ann.\ Rev.\ Nucl.\ Part.\ Sci.\  {\bf 50}, 525 (2000)
[arXiv:hep-ph/0007218].

\bibitem{ref:kkst} For a study of $A_{TT}$ at small transverse momentum
of the Drell-Yan photon that includes all-order soft-gluon resummations, 
see: H.~Kawamura, J.~Kodaira, H.~Shimizu and K.~Tanaka,
contribution to the ``7th DESY Workshop on Elementary Particle Theory: 
Loops and Legs in Quantum Field Theory'', Zinnowitz, Germany, April 2004,
Nucl.\ Phys.\ Proc.\ Suppl.\  {\bf 135}, 19 (2004).

\bibitem{dypax1} P.~Lenisa and F.~Rathmann  [the PAX Collaboration],
arXiv:hep-ex/0505054 and {\tt http://www.fz-juelich.de/ikp/pax/};
see also: F.~Rathmann and P.~Lenisa, contribution to the ``16th International 
Spin Physics Symposium (SPIN 2004)'', Trieste, Italy, October 2004,
arXiv:hep-ex/0412078; P.~Lenisa {\it et al.}, contribution to the
``2nd High-Energy Physics Conference in Madagascar (HEP-MAD 04)'', 
Antananarivo, Madagascar, September 2004, eConf {\bf C0409272}, 014 (2004)
[arXiv:hep-ex/0412063].

\bibitem{dypax1a} GSI-ASSIA Technical Proposal, Spokesperson: R. Bertini,
{\tt http://www.gsi.de/documents/DOC-2004-Jan-152-1.ps};
M.~Maggiora  [the ASSIA Collaboration], arXiv:hep-ex/0504011.

\bibitem{dypax2} 
M.~Anselmino, V.~Barone, A.~Drago and N.~N.~Nikolaev,
Phys.\ Lett.\ B {\bf 594}, 97 (2004) [arXiv:hep-ph/0403114].

\bibitem{dypax3} A.~V.~Efremov, K.~Goeke and P.~Schweitzer,
Eur.\ Phys.\ J.\ C {\bf 35}, 207 (2004) [arXiv:hep-ph/0403124].

\bibitem{dypax4} For recent Monte-Carlo studies of the Drell-Yan 
cross section and spin asymmetries at GSI energies, see: 
A.~Bianconi and M.~Radici, Phys.\ Rev.\ D {\bf 71}, 
074014 (2005) [arXiv:hep-ph/0412368]; arXiv:hep-ph/0504261.

\bibitem{JG}  J.~w.~Qiu and G.~Sterman, Nucl.\ Phys.\ B {\bf 353}, 105 (1991);
Nucl.\ Phys.\ B {\bf 353}, 137 (1991).

\bibitem{dyresum1} G.~Sterman, Nucl.\ Phys.\ B {\bf 281}, 310 (1987).

\bibitem{dyresum2} S.~Catani and L.~Trentadue, Nucl.\ Phys.\ B {\bf 327}, 
323 (1989); Nucl.\ Phys.\ B {\bf 353}, 183 (1991).

\bibitem{Kidonakis} N.~Kidonakis,
Int.\ J.\ Mod.\ Phys.\ A {\bf 15}, 1245 (2000) [arXiv:hep-ph/9902484].

\bibitem{IJ} A.~V.~Manohar,
Phys.\ Rev.\ D {\bf 68}, 114019 (2003) [arXiv:hep-ph/0309176];
A.~Idilbi and X.~Ji, arXiv:hep-ph/0501006.

\bibitem{SCET} C.~W.~Bauer, S.~Fleming, D.~Pirjol and I.~W.~Stewart,
Phys.\ Rev.\ D {\bf 63}, 114020 (2001) [arXiv:hep-ph/0011336];
C.~W.~Bauer, S.~Fleming and M.~E.~Luke,
Phys.\ Rev.\ D {\bf 63}, 014006 (2001) [arXiv:hep-ph/0005275].

\bibitem{renormalon} G.~'t Hooft, Nucl.\ Phys.\ B {\bf 138}, 1 (1978);
A.~H.~Mueller, Nucl.\ Phys.\ B {\bf 250}, 327 (1985); 
M.~Beneke and V.~M.~Braun, arXiv:hep-ph/0010208, and references
therein; 
Y.~L.~Dokshitzer, G.~Marchesini and B.~R.~Webber,
Nucl.\ Phys.\ B {\bf 469}, 93 (1996) [arXiv:hep-ph/9512336];
JHEP {\bf 9907}, 012 (1999) [arXiv:hep-ph/9905339]; 
G.~P.~Korchemsky and G.~Sterman, Nucl.\ Phys.\ B {\bf
437}, 415 (1995) [arXiv:hep-ph/9411211];
R.~Akhoury and V.~I.~Zakharov, Phys.\ Lett.\ B {\bf 357}, 646 (1995)
[arXiv:hep-ph/9504248]; Nucl.\ Phys.\ B {\bf 465}, 295 (1996)
[arXiv:hep-ph/9507253]; Phys.\ Rev.\ Lett.\  {\bf 76}, 2238 (1996)
[arXiv:hep-ph/9512433].

\bibitem{cspv} H.~Contopanagos and G.~Sterman, Nucl.\ Phys.\ B {\bf 419}, 
77 (1994) [arXiv:hep-ph/9310313]; M.~Beneke and V.~M.~Braun,
Nucl.\ Phys.\ B {\bf 454}, 253 (1995) [arXiv:hep-ph/9506452];
see also: M.~Beneke, Phys.\ Rept.\  {\bf 317}, 1 (1999)
[arXiv:hep-ph/9807443]; 
G.~Sterman and W.\ Vogelsang, arXiv:hep-ph/9910371; 
Phys.\ Rev.\ D {\bf 71}, 014013 (2005) [arXiv:hep-ph/0409234].

\bibitem{KSV} A.~Kulesza, G.~Sterman and W.~Vogelsang,
Phys.\ Rev.\ D {\bf 66}, 014011 (2002) [arXiv:hep-ph/0202251];
Phys.\ Rev.\ D {\bf 69}, 014012 (2004) [arXiv:hep-ph/0309264].

\bibitem{ref:dy2}
W.\ Vogelsang,  Phys.\ Rev.\ D {\bf 57}, 1886 (1998)
[arXiv:hep-ph/9706511].

\bibitem{vogtnnll} A.~Vogt,
Phys.\ Lett.\ B {\bf 497}, 228 (2001) [arXiv:hep-ph/0010146].

\bibitem{KT} J.~Kodaira and L.~Trentadue, 
Phys.\ Lett.\ B {\bf 112}, 66 (1982); Phys.\ Lett.\ B {\bf 123}, 
335 (1983);\\ S.~Catani, E.~D'Emilio and L.~Trentadue,
Phys.\ Lett.\ B {\bf 211}, 335 (1988).

\bibitem{Eynck:2003fn} T.~O.~Eynck, E.~Laenen and L.~Magnea,
JHEP {\bf 0306}, 057 (2003) [arXiv:hep-ph/0305179].

\bibitem{Catani:1996yz} S.~Catani, M.~L.~Mangano, P.~Nason
and L.~Trentadue, Nucl.\ Phys.\ B {\bf 478}, 273 (1996)
[arXiv:hep-ph/9604351].

\bibitem{CMN} S.~Catani, M.~L.~Mangano and P.~Nason,
JHEP {\bf 9807}, 024 (1998) [arXiv:hep-ph/9806484].

\bibitem{cat} S.~Catani, D.~de~Florian and M.~Grazzini, in:
W.~Giele {\it et al.}, arXiv:hep-ph/0204316; \\
S.~Catani, D.~de Florian and M.~Grazzini,
JHEP {\bf 0201}, 015 (2002)  [arXiv:hep-ph/0111164].

\bibitem{Kra98} M.\ Kramer, E.\ Laenen and M.\ Spira, 
 Nucl.\ Phys.\ B{\bf 511}, 523 (1998)
[arXiv:hep-ph/9611272].

\bibitem{thrrap}  G.~Sterman and W.~Vogelsang,
JHEP {\bf 0102}, 016 (2001) [arXiv:hep-ph/0011289].

\bibitem{grv} M.\ Gl\"{u}ck, E.\ Reya, and A.\ Vogt,
Eur.\ Phys.\ J.\ {\bf C5}, 461 (1998) [arXiv:hep-ph/9806404].

\bibitem{vN} R.~Hamberg, W.~L.~van Neerven and T.~Matsuura,
Nucl.\ Phys.\ B {\bf 359}, 343 (1991) [Erratum-ibid.\ B {\bf 644}, 403 (2002)].
For the numerical evaluation of the NNLO corrections we used the code
provided in that reference. \\ P.~J.~Rijken and W.~L.~van Neerven,
Phys.\ Rev.\ D {\bf 51}, 44 (1995) [arXiv:hep-ph/9408366]; \\
see also: R.~V.~Harlander and W.~B.~Kilgore,
Phys.\ Rev.\ Lett.\  {\bf 88}, 201801 (2002) [arXiv:hep-ph/0201206].

\bibitem{mrstnnlo} A.~D.~Martin, R.~G.~Roberts, W.~J.~Stirling 
and R.~S.~Thorne, Phys.\ Lett.\ B {\bf 604}, 61 (2004)
[arXiv:hep-ph/0410230]; see also:
J.~Bl\"{u}mlein, H.~B\"{o}ttcher and A.~Guffanti,
Nucl.\ Phys.\ Proc.\ Suppl.\  {\bf 135}, 152 (2004)
[arXiv:hep-ph/0407089].

\bibitem{3loop} S.~Moch, J.~A.~M.~Vermaseren and A.~Vogt,
Nucl.\ Phys.\ B {\bf 688}, 101 (2004) [arXiv:hep-ph/0403192];
{\it ibid.} {\bf 691}, 129 (2004) [arXiv:hep-ph/0404111].

\bibitem{dyrev} for reviews of Drell-Yan data, see: 
K.~Freudenreich, Int.\ J.\ Mod.\ Phys.\ A {\bf 5} (1990) 3643;
W.~J.~Stirling and M.~R.~Whalley, J.\ Phys.\ G {\bf 19} (1993) D1.

\bibitem{wa39} M.~Corden {\it et al.} [WA39 Collaboration], 
Phys.\ Lett.\ B {\bf 96}, 417 (1980).

\bibitem{grvpi} M.~Gl\"{u}ck, E.~Reya and A.~Vogt, 
Z.\ Phys.\ C {\bf 53}, 651 (1992).

\bibitem{ref:soffer}  J.~Soffer, Phys.\ Rev.\ Lett.\  {\bf 74}, 1292 (1995) 
[arXiv:hep-ph/9409254]; D.~W.~Sivers, Phys.\ Rev.\ D {\bf 51} (1995) 4880.

\bibitem{grsv} M.~Gl\"{u}ck, E.~Reya, M.~Stratmann and W.~Vogelsang,
Phys.\ Rev.\ D {\bf 63}, 094005 (2001)  [arXiv:hep-ph/0011215].

\bibitem{ref:nlokernels} S.~Kumano and M.~Miyama,
Phys.\ Rev.\ D {\bf 56}, 2504 (1997)  [arXiv:hep-ph/9706420];
A.~Hayashigaki, Y.~Kanazawa and Y.~Koike,
Phys.\ Rev.\ D {\bf 56}, 7350 (1997)  [arXiv:hep-ph/9707208].

\bibitem{ratcl04} P.~G.~Ratcliffe, arXiv:hep-ph/0412157.

\bibitem{stwvpdf} G.~Sterman and W.~Vogelsang, in: Proceedings of 
``Physics at Run II: QCD and Weak Boson Physics Workshop'', Fermilab,
1999, arXiv:hep-ph/0002132.

\bibitem{largexdis} S.~Simula,
Phys.\ Lett.\ B {\bf 493}, 325 (2000) [arXiv:hep-ph/0005315];
S.~Schaefer, A.~Schafer and M.~Stratmann,
Phys.\ Lett.\ B {\bf 514}, 284 (2001) [arXiv:hep-ph/0105174];
E.~Gardi and R.~G.~Roberts,
Nucl.\ Phys.\ B {\bf 653}, 227 (2003) [arXiv:hep-ph/0210429].

\end{thebibliography}
\end{document}